\documentclass[prd,aps,floats,twocolumn,nofootinbib]{revtex4-1}
\usepackage{slashed}
\usepackage{mathtools}
\usepackage{amsfonts}
\usepackage{amssymb}
\usepackage{epsfig}
\usepackage{rotating}
\usepackage{url}
\usepackage{times}
\usepackage{color}
\usepackage{bm}
\usepackage{xcolor,colortbl}
\usepackage{hyperref}
\usepackage[alsoload=hep]{siunitx}
\usepackage{enumitem}
\usepackage{fancyhdr}
\usepackage{anyfontsize}
\usepackage{wasysym}

\newcommand{\grad}{\ensuremath{\vec{\nabla}}}

\newcommand{\sign}{\ensuremath{{\rm Sign}}}

\newcommand{\signY}{\sigma_Y}

\newcommand{\nn}{\nonumber}
\newcommand{\ph}{\phantom}

\newcommand{\dsdt}{\frac{ds}{dt}}

\newcommand{\dphidt}{\frac{d\phi}{dt}}
\newcommand{\appropto}{\mathrel{\vcenter{
			\offinterlineskip\halign{\hfil$##$\cr
				\propto\cr\noalign{\kern2pt}\sim\cr\noalign{\kern-2pt}}}}}

\newcommand{\muh}{\ensuremath{\hat{\mu}}}

\newcommand{\metE}{\ensuremath{\tilde{g}}}
\newcommand{\LambdaE}{\ensuremath{\tilde{\Lambda}}}

\newcommand{\HE}{\ensuremath{\tilde{H}}}
\newcommand{\NE}{\ensuremath{\tilde{N}}}

\newcommand{\kgal}{\ensuremath{k_{\rm gal} }}
\newcommand{\epsilonk}{\ensuremath{\epsilon_{\rm c} }}

\begin{document}

\title{Cosmology of the Galileon extension of Bekenstein's theory of relativistic Modified Newtonian Dynamics}

\author{T.G.~Z\l o\'{s}nik$^{1}$}
\email{zlosnik@fzu.cz}
\author{C.~Skordis$^{1,2}$}
\email{skordis@ucy.ac.cy}

\affiliation{
	$^1$ Institute of Physics of the Czech Academy of Sciences, Na Slovance 1999/2, 182 21, Prague \\
	$^2$ Department of Physics, University of Cyprus, 1, Panepistimiou Street, 2109, Aglantzia, Cyprus 
}

\begin{abstract}
	A generalization of Bekenstein's Tensor-Vector-Scalar (TeVeS) model of modified gravity has recently been proposed as 
an alternative to dark matter. This model -- which we will refer to as g-TeVeS -- utilizes a Galileon-induced Vainshtein 
mechanism to suppress modifications to General Relativity in strong gravity regimes and so avoids the need to introduce the 
baroque kinetic terms that typically exist in relativistic models of Modified Newtonian Dynamics (MOND).
	
	We explore the behavior of this model in spacetimes with exact Friedmann-Robertson-Walker (FRW) symmetry. The 
ability of the theory to recover MOND phenomenology places restrictions on the theory's parameter space and it is found 
that within an estimate of this area of parameter space the theory cannot successfully approximate the Friedmannian 
cosmological behavior of cold dark matter. It is found that much closer agreement may be recovered in other regions of 
the theory's parameter space and the reasons for this are discussed.  
\end{abstract}

\maketitle
\subsection{Introduction}
\label{Section_Introduction}

Milgrom's observation \cite{Milgrom1983} that a wide variety of the astrophysical phenomena usually attributed to the effects of dark matter can instead be accounted for by a modification to the dynamics of visible matter has provided an intriguing hint that something may be missing in our understanding of gravity and/or inertia. In its original formulation, Milgrom's Modified Newtonian Dynamics (MOND) was non-relativistic in the same sense that Newton's theory of gravity is.  If Newtonian gravity is a limiting form of General Relativity, what is MOND a limiting form of? One formulation of MOND is as a modified Poisson equation:

\begin{equation}
\grad\cdot\left[ \mu_m(x) \grad\Phi\right] = 4\pi G_{N} \rho_b
\label{modpoisson}
\end{equation}
where $\Phi$ is the gravitational potential felt by non-relativistic test particles, 
  $x=|\grad\Phi|/a_0$,  $\rho_b$ is the density 
of baryonic matter, $G_{N}$ is the locally measured value of Newton's gravitational constant, $a_{0}$ is a constant 
with the dimensions of acceleration and $\mu_m(x)$ is a function subject to the limiting forms $\mu_m \rightarrow 1$ 
as $x\gg 1$ and $\mu_m \rightarrow x$ as $x \ll 1$ but is otherwise unspecified; an explicit form such as $\mu_m(x)=x/(1+x)$ 
has usually been chosen but such forms lack theoretical motivation \footnote{There do, however, exist examples where 
non-relativistic MOND phenomenology is recovered with a specific counterpart of $\mu_m(x)$, whose form is 
fixed by heuristic theoretical considerations \cite{Milgrom1998,KlinkhamerKopp2011,PazyArgaman2011}.}.

A significant amount of research has gone into studying the consequences of (\ref{modpoisson}) in astrophysical systems 
\cite{SandersMcGaugh2002,BekensteinMagueijo2006,GentileEtAl2007,WuEtAl2009,DaiStarkman2010,ZhaoFamaey2010,MagueijoMozaffari2011,
FamaeyMcGaugh2011,Heesetal2015,MargalitShaviv2015,PereiraOverduinPoyneer2016,Ko2016}; however, the lack of a fully-relativistic 
formulation of the theory makes it difficult to know the realm of the equation's validity. It was also attempted to
derive the MOND formula from fundamental theory~\cite{Verlinde2016}. A number of relativistic 
theories that recover MOND-like phenomenology have been proposed \cite{BekensteinMilgrom1984,Bekenstein2004,NavarroVanAcoleyen2005,
Sanders2007,ZlosnikFerreiraStarkman2006,Milgrom2009,BlanchetMarsat2011,DeffayetEsposito-FareseWoodard2011,Woodard2014,Khoury2014,KimEtAl2016}. 
All of these examples possess a 
similar ambiguity to (\ref{modpoisson}) in that all possess a function in the Lagrangian that must be chosen by hand. This makes it 
difficult to know exactly what a given theory predicts. Each of the examples involve the introduction of new degrees of freedom into 
physics and it can be that the line between their interpretation as an additional `dark force' in nature or simply a type of dark 
matter becomes blurred. Indeed, alternatively there have been a number of attempts to produce results for $\Phi$ similar to that in 
solutions of (\ref{modpoisson}) by instead evoking a dark matter with exotic dynamics and coupling to matter 
\cite{BlanchetLeTiec2008,BerezhianiKhoury20151,BerezhianiKhoury20152}.

Recently a model that produces MOND-like phenomenology whilst avoiding the use of an unspecified function has been proposed by Babichev, Deffayet, and Esposito-Farese \cite{BabichevDeffayetEsposito-Farese2011}. This theory uses Bekenstein's TeVeS theory as basis~\cite{Bekenstein2004,Skordis2009}
but extends it with the addition of a Galileon-type term~\cite{DeffayetEtAl2011} and the removal of the free function.
The Galileon term leads to the Vainshtein screening~\cite{Vainshtein1972}  
 of the force generated by the scalar field  around the high-curvature environment of the 
solar system.  We will therefore refer to the current model as g-TeVeS for Galileon extended Tensor-Vector-Scalar theory.
We also note that applying the use of screening mechanisms to MOND has also been investigate recently in~\cite{BurrageCopelandMillington2016} 
by utilizing the symmetron mechanism~\cite{HinterbichlerKhoury2010}.

The action for this theory is as follows:
\begin{widetext}
\begin{align}
S &= S_{{\rm grav}}[\tilde{g}_{\mu\nu}]+S_{A}[\tilde{g}_{\mu\nu},A_{\mu},\lambda]+S_{\phi}[\tilde{g}_{\mu\nu},\phi]
 +S_{mat}[g_{\mu\nu},\chi] \label{kmouflage_action}
\end{align}
\begin{align}
S_{{\rm grav}} &= \frac{1}{16\pi G}\int d^{4}x \sqrt{-\tilde{g}} \left(\tilde{R}-2\tilde{\Lambda}\right)
\label{gravityaction}
\\
S_{A} &= -\frac{1}{16\pi G}\int d^{4}x \sqrt{-\tilde{g}}  
 \left[K^{\alpha\beta\mu\nu}\tilde{\nabla}_{\alpha}A_{\mu}\tilde{\nabla}_{\beta}A_{\nu}+\lambda(\tilde{g}^{\mu\nu}A_{\mu}A_{\nu}+1)\right]
\label{aetheraction}
\\
S_{\phi} &= -\frac{1}{8\pi G}\int d^{4}x \sqrt{-\tilde{g}} 
\left(\epsilonk\tilde{g}^{\mu\nu}\tilde{\nabla}_{\mu}\phi\tilde{\nabla}_{\nu}\phi+\frac{2}{3\tilde{a}_{0}}\tilde{g}^{\mu\nu}\tilde{\nabla}_{\mu}\phi\tilde{\nabla}_{\nu}\phi\sqrt{|\tilde{g}^{\alpha\beta}\tilde{\nabla}_{\alpha}\phi\tilde{\nabla}_{\beta}\phi|}\right. 
\nn 
\\
&\left. +\frac{2\kgal}{3}\tilde{\epsilon}^{\alpha\beta\gamma\delta}\tilde{\epsilon}^{\mu\nu\rho\sigma}
\tilde{\nabla}_{\alpha}\phi\tilde{\nabla}_{\mu}\phi\tilde{\nabla}_{\nu}\tilde{\nabla}_{\beta}\phi \tilde{R}_{\gamma\delta\rho\sigma}
+m_\phi^2 \phi^2  \right)
\label{phiaction}
\end{align}
\end{widetext}
where $\tilde{R}$ is the Ricci-scalar corresponding to the connection $\tilde{\nabla}_{\mu}$ compatible with the metric $\tilde{g}_{\mu\nu}$
(i.e. $\tilde{\nabla}_{\mu}\tilde{g}_{\alpha\beta}=0$), $\tilde{g}$ is the determinant of $\tilde{g}_{\mu\nu}$ and $\tilde{\epsilon}^{\alpha\beta\gamma\delta}$ is the completely antisymmetric tensor built using it, and $\chi$ represent the matter fields in the universe, 

\begin{align}
K^{\alpha\beta\mu\nu} &\equiv  \left(\right. c_{1}\tilde{g}^{\alpha\beta}\tilde{g}^{\mu\nu}+c_{2}\tilde{g}^{\alpha\mu}\tilde{g}^{\beta\nu}\nn\\
&+c_{3}\tilde{g}^{\alpha\nu}\tilde{g}^{\beta\mu}+c_{4}A^{\alpha}A^{\beta}\tilde{g}^{\mu\nu}\left. \right)
\end{align}
where $A^{\mu}\equiv \tilde{g}^{\mu\nu}A_{\nu}$, 
and matter is taken to couple to the metric $g_{\alpha\beta}$:

\begin{equation}
g_{\mu\nu} \equiv e^{-2\phi}\tilde{g}_{\mu\nu}-2\mathrm{sinh}(2\phi)A_{\mu}A_{\nu} 
\label{twometrics}
\end{equation}
By comparison, the action of Bekenstein's TeVeS theory differs only in the form of $S_{\phi}$ \cite{Bekenstein2004}. In TeVeS there is no Galileon term and, in its `diagonal frame' formulation 
\cite{Skordis2009}, the equivalent action to $S_{\phi}$ takes the form:

\begin{equation}
S_{\phi}^{{\rm (TeVeS)}} =  -\frac{1}{8\pi G}\int d^{4}x\sqrt{-\tilde{g}} f(X)
\end{equation}
where $X\equiv \tilde{g}^{\mu\nu}\tilde{\nabla}_{\mu}\phi\tilde{\nabla}_{\nu}\phi$ and- as in the case of the function $\mu_{m}$ in (\ref{modpoisson})- the function $f$ is required to take certain limiting forms but it is otherwise unspecified; that there are no free functions present in (\ref{phiaction}) is what represents a simplification over TeVeS.

We now discuss the individual terms in (\ref{kmouflage_action}) in more detail. The first term, $S_{{\rm grav}}$ is simply the Einstein-Hilbert term for the metric $\tilde{g}_{\mu\nu}$ along with a cosmological constant term. One may additionally consider a cosmological term with respect to the metric $g_{\mu\nu}$:

\begin{equation}
S_{\Lambda} = -\frac{1}{8\pi G}\int  d^{4}x \sqrt{-g}\Lambda
\end{equation}
The action $S_{A}$ is the Einstein-Aether action \cite{JacobsonMattingly2000}. This is the most general action for a fixed-norm timelike vector field coupled to $\tilde{g}_{\mu\nu}$ which produces field equations second-order in time. The first term in $S_{\phi}$, proportional to the constant $\epsilonk$ is a canonical kinetic term; it is included as a step towards providing a well-defined Cauchy problem for the theory in the limit of weak fields
\cite{Bruneton2006}. The second term, proportional to the constant $1/\tilde{a}_{0}$ is the term that provides an analog to the limit where $\mu_m(x)\sim x$ in (\ref{modpoisson}); it is this limit which is responsible for a dark matter-like effect in astrophysical systems. The third term is is an example of the `Paul' Galileon Lagrangian of the Fab Four scalar field models \cite{Charmousisetal2011}. Its role is to suppress the additional force on test bodies in high curvature regions that would otherwise exist due to the $1/\tilde{a}_{0}$ term. These effects will be discussed in more detail in Section \ref{Section_Parameterspace}. Finally, the term proportional $m_{\phi}^{2}$ is a mass term for the scalar field.

\subsection{The parameter space}
\label{Section_Parameterspace}
We now briefly discuss the restrictions on the parameters of the g-TeVeS theory that were derived by the authors of \cite{BabichevDeffayetEsposito-Farese2011}. It is found in the theory that on galactic and sub-galactic scales, the non-relativistic acceleration $\mathbf{g}$ felt by a test body obeys the relation

\begin{equation}
\mathbf{g} =  - \left(\grad\Phi_{N} + \grad\phi\right)
\end{equation}
where $\grad$ is the spatial gradient operator and $\Phi_{N}$ is the Newtonian potential due to baryonic matter; thus, spatial gradients of the field $\phi$ provide an extra force acting on test bodies and the model is constructed to make that force resemble the phenomenology implied by MOND.
It is found in \cite{BabichevDeffayetEsposito-Farese2011} that in spherical symmetry, outside of a gravitating object with mass $M$, $d\phi/dr$ takes the form:

\begin{widetext}
\begin{equation}
\frac{d\phi}{dr} =  \left(\sqrt{\frac{8\kgal}{r^{2}}+\frac{r^{2}}{G_{N}Ma_{0}}+\left(\frac{\epsilonk r^{2}}{2G_{N}M}\right)^{2}}+\frac{\epsilonk r^{2}}{2G_{N}M}\right)^{-1} 
\label{dphidr}
\end{equation}
\end{widetext}
Importantly, it has been assumed that the following equalities hold:

\begin{align}
G  &= G_{N}\quad, \quad \tilde{a}_{0} = a_{0} 
\label{gandao}
\end{align}
i.e. that the parameters $\{G,\tilde{a}_{0}\}$ that appear in the action (\ref{kmouflage_action}) are to be identified respectively with the locally measured value of Newton's constant $G_{N}$ and the scale $a_{0}$ appearing in MOND phenomenology. This is a non-trivial assumption. It is known in the case of TeVeS that $G_{N}$ and $a_{0}$ will generally depend on a number of additional quantities such as the `background' value of the scalar field $\phi$ (in the case of $a_{0}$), certain constants in the tensor $K^{\alpha\beta\mu\nu}$ should they be non-zero, and constants in the scalar field action \cite{Bekenstein2004,BekensteinSagi2008,Skordis2009}.
In this work we also assume that the equalities (\ref{gandao}) hold though a more comprehensive analysis
involving an extended version of the Parameterized Post-Newtonian Vanshteinian (PPNV) formalism~\cite{AvilezEtAl2015} may 
 show otherwise (as is the case of TeVeS).

The authors of \cite{BabichevDeffayetEsposito-Farese2011} identify two distinct length scales 
$r_{V}$ and $r_{M}$ $\gg r_{V}$ which mark transitions between forms of $d\phi/dr$:
\begin{widetext}
\begin{align}
r_{V} &= (8\kgal G_{N}M a_{0})^{\frac{1}{4}} \quad\quad \left(r \ll r_{V} \rightarrow \frac{d\phi}{dr}\sim\frac{r}{\sqrt{8\kgal}}\right)
 \\
r_{M} &= \frac{1}{\epsilonk}\sqrt{\frac{G_{N}M}{a_{0}}} \quad\quad \left(r_{V} \ll r \ll r_{M} \rightarrow \frac{d\phi}{dr}\sim\frac{\sqrt{G_{N}M a_{0}}}{r} ; \,\, r \gg r_{M}\rightarrow \frac{d\phi}{dr} \sim \frac{G_{N}M}{\epsilonk r^2}\right)
\end{align}
\end{widetext}

The parameters should be such that for regions in the vicinity of gravitating sources where there is little evidence for dark matter, then $|\grad\phi|/|\grad\Phi_{N}|\ll 1$. For example, when considering the gravity due to the sun, if scales $r$ within the solar system are less than $r_{V}$ then the additional force due to $\phi$ is proportional to $r$ and actually decreases the closer one gets to the sun. This is the Vainshtein mechanism created by the presence of the Galileon term.

It is required that as one moves to larger distances from a gravitating source then one reaches scales $r > r_{V}$ 
 (e.g. when considering the motion of stars towards the outer regions of galaxies) then the additional acceleration due to $\phi$ is approximately $-\sqrt{G_{N}M a_{0}}/r$ where

\begin{equation}
a_{0} \sim 1.2 \times 10^{-10} m/s^{2} \sim cH_{0}/6
\end{equation}
and $H_{0}$ is the measured value of the Hubble constant today.
However, this acceleration cannot persist on much smaller astrophysical scales such as those of the Solar System as its presence would result in effects on planetary orbits that are not observed \cite{Sanders2006}. If, however, the parameter $\kgal$ is non-zero and typical planetary distances from the sun are significantly smaller than the radius $r_{V}$ then the additional acceleration provided by $\phi$ will be of magnitude $r/\sqrt{8\kgal}$. Thus, if $\kgal$ is sufficiently big then the additional acceleration experienced by objects in the solar system will be currently undetectable.

We assign a rough measure $r_{SOL}$ of the maximum distance from the sun probed by experiment to be of the order of lengthscales associated with Pluto's orbit: 

\begin{equation}
r_{SOL} \sim 10^{9}G_{N}M_{\astrosun}
\end{equation}
and we require that $r_{V}$ for the sun is assumed greater than $r_{SOL}$ so 
that the anomalous acceleration is of the form $r/\sqrt{8\kgal}$ for $r<r_{SOL}$. This gives the condition:

\begin{equation}
\kgal \gtrsim \frac{10^{36}(G_{N}M_{\astrosun})^{3}}{8 a_{0}}
\end{equation}
However, as the ratio $r_{V}/r_{M}$ grows as $\kgal^{1/4}$, there is an upper bound on how big $\kgal$ can be. It is also observed that systems with baryonic mass as little as $10^{3}M_{\astrosun}$ display mass discrepancy effects so the $r_{V}$ should be smaller than $r_{M}$ for these systems:

\begin{equation}
\kgal< \frac{10^{3}G_{N}M_{\astrosun}}{8 a_{0}^{3}}
\end{equation}
Hence the constant $\kgal$ is restricted from both sides:

\begin{equation}
\left(2\times 10^{-9}\frac{1}{a_{0}}\right)^{4} \lesssim \kgal < \left(4\times 10^{-6}\frac{1}{a_{0}}\right)^{4}
\end{equation}
The effects of an acceleration dominated by the $1/r$ term have been observed at distances from the gravitating source corresponding to at least $10r_{M}$. In the g-TeVeS model, the MOND regime is exited for $r>r_{M}/\epsilonk$ and so we have:

\begin{equation}
\epsilonk < 0.1
\end{equation}
In summary then, the value $\tilde{a}_{0}$ is essentially fixed by accounting for the motion of stars within galaxies which display evidence for dark matter; the value $\kgal$ is restricted to be small enough so that the Galileon term does not dominate in the rather low baryonic mass systems which still display evidence for dark matter but that nonetheless is big enough to dominate in the Solar System; the value $\epsilonk$ is restricted to be small enough so that the MOND term dominates for a sufficiently great span of scales where there is evidence for dark matter. 

Additionally there are the four constants $\{c_{i}\}$ in (\ref{aetheraction}). In the canonical Einstein-Aether theory (the theory which is a limit of (\ref{kmouflage_action}) as $\phi\rightarrow \mathrm{cst.}=0$), it is known that in the quasistatic weak-field regime that the parameters $c_{1}$ and $c_{4}$  contribute to a modification of the relation between $G_{N}$ and $G$; these and other parameters are involved in deviations of the theory's parameterized post-Newtonian (PPN) parameters from those of General Relativity \cite{FosterJacobson2005}. It likely must be assumed then that the $\{c_{i}\}$ in general take values such that they have a negligible effect on PPN parameters, including the ratio $G_{N}/G$. Once again,
this would involve an extended version of the PPNV formalism~\cite{AvilezEtAl2015} and is left for a future investigation.

A comprehensive comparison between g-TeVeS and the standard $\Lambda$-cold dark matter cosmological model ($\Lambda\mathrm{CDM}$) would require development of its cosmological perturbation theory and the computation of its predictions for the growth of cosmic structure. In the present paper we take the first steps in this direction and examine how close the expansion history of the universe in g-TeVeS can be to that of $\Lambda\mathrm{CDM}$. We will see whether there exist parameters in the restricted parameter space discussed in this section which lead to similarity to $\Lambda\mathrm{CDM}$ and we will also explore the remainder of the parameter space, it is important to do so because- as argued- it is not yet clear what restrictions the astrophysical limit of the theory places on parameters. 

\subsection{FRW Symmetry}
\label{Section_FRW}
We now derive the field equations of the g-TeVeS model  assuming FRW symmetry. In this symmetry the Einstein metric $\tilde{g}_{\mu\nu}$ can be cast into the following form:
\begin{equation}
	\tilde{g}_{\mu\nu}dx^{\mu}\otimes dx^{\nu} = -\tilde{N}^{2}(t)dt^{2}+b^{2}(t)\gamma_{ij}dx^{i}\otimes dx^{j}
\end{equation}
where one can further adopt the coordinates
\begin{equation}
	\gamma_{ij}dx^{i}\otimes dx^{j} = \frac{1}{1-\kappa r^{2}}dr^{2}+r^{2}dS^2_{(2)}
\end{equation}
where $\kappa$ is the spatial curvature scalar of co-moving spatial hypersurfaces and $dS^2_{(2)}$ the line element of a $2$-sphere.

The field $A_{\mu}$, is unit-timelike with respect to the metric $\tilde{g}_{\mu\nu}$ due to the Lagrangian constraint in the action (\ref{aetheraction}). In FRW symmetry this is satisfied by the following ansatz, which we will adopt:
\begin{equation}
A_{\mu} =  \mp \tilde{N} \delta^{0}_{\ph{0}\mu}.
\end{equation}
Without loss of generality we will choose the minus sign option, so that the vector $A^\mu \equiv \tilde{g}^{\mu\nu}A_{\nu}$ `points' in the future direction. The scalar field $\phi$ is assumed to depend only on the coordinate $t$. The matter frame metric $g_{\mu\nu}$ then takes the form
\begin{equation}
	g_{\mu\nu}dx^{\mu}\otimes dx^{\nu} = -N^{2}(t)dt^{2}+a^{2}(t)\gamma_{ij}dx^{i}\otimes dx^{j}
\end{equation}
where
\begin{align}
	N &= \tilde{N}e^{\phi} 
\\
	a &= b e^{-\phi}
\end{align}
We define new variables in order to perform a Hamiltonian formulation of the action.
Rather than the scale factor $b$ we will work with a variable $s$ defined by 
\begin{equation}
	s = \ln b
\end{equation}
assuming that  $b$ does not pass through zero.

We furthermore introduce the variables:
\begin{align}
	\tilde{h} &\equiv \frac{ds}{dt}
\label{k_mouflage_HE_diff}
\\
	y &\equiv \frac{d\phi}{dt}
\label{k_mouflage_Z_diff}
\end{align}
and enforce the constraints via Lagrangian multiplier terms, $\pi_s$ and $\pi_\phi$. Using some useful results provided in Appendix \ref{Appendix_FRW}, the FRW-reduced action can be shown to be:
\begin{widetext}
\begin{align}
 S &= \frac{1}{8\pi G}\int dt \bigg\{
-  \frac{3K_F b^3}{\NE}  \tilde{h}^2  +  3\kappa \NE b
-  \NE b^3  \LambdaE
+ b^3\left( \frac{\epsilonk}{\NE} y^2 + \frac{2}{3\tilde{a}_{0}\NE^2} |y|^3 - \NE m_\phi^2 \phi^2  \right)
\nonumber
\\
&
+ 8\kgal \frac{b}{\NE^3  } y^3 \tilde{h} \left( \kappa +  \frac{b^2\tilde{h}^2}{\NE^2}  \right) 
+ \pi_s\left( \dsdt - \tilde{h}\right) 
+ \pi_\phi\left( \dphidt -y\right) 
\bigg\}
+ S_m \label{FRWaction}
\end{align}
\end{widetext}
where following~\cite{Skordis2008} we define $K_{F}\equiv 1+(c_{1}+3c_{2}+c_{3})/2$.
A way to explicitly include matter is by considering the definition of the matter action in terms of its variation which leads to the stress-energy tensor; in particular we have that
\begin{widetext}
\begin{align}
\delta S_m &=  -\frac{1}{2} \int d^{4}x \sqrt{-g} T_{\mu\nu}\delta g^{\mu\nu} 
\nn 
\\
&= -\frac{1}{2} \int  d^{4}x \sqrt{-g}
\bigg[ 
\left(e^{2\phi} T_{\mu\nu} + 4\sinh(2\phi) A^\alpha A_\mu  T_{\alpha\nu} \right)  \delta \metE^{\mu\nu} 
+  2 \left( T_{\mu\nu}\tilde{g}^{\mu\nu} + 2 e^{-2\phi}  T_{\mu\nu} A^\mu A^\nu  \right) \delta \phi\nn \\
& + 4\sinh(2\phi)  T_{\mu\nu}  A^\mu \metE^{\nu \alpha}  \delta A_\alpha
\bigg]
\label{general_fluid_variation_TeVeS}
\end{align}
\end{widetext}

We take the matter content to be described a perfect fluid, where for each matter species $T_{\mu\nu}= \rho u_{\mu}u_{\nu}+ P(g_{\mu\nu}+u_{\mu}u_{\nu})$ where $u_{\mu}\equiv g_{\mu\nu}u^{\nu}$ and 
$g_{\mu\nu}u^{\mu}u^{\nu}=-1$. After integration over the trivial spatial integral in the action yields the FRW-reduced form
\begin{align}
\delta S_m &= -\int dt  e^{-2\phi} b^3
\bigg[
\rho \delta  \NE - \frac{3\NE }{b}     P \delta b 
\nn
\\
&
+
  \NE \left( \rho + 3P \right) \delta \phi
\bigg]
\label{FRW_fluid_variation_TeVeS}
\end{align}
where we have assumed that $u^{\mu}$ aligns with the direction of cosmic time basis vector $(\partial_{t})^{\mu}$ i.e. $u^\mu= (\frac{1}{N},0,0,0) = (\frac{e^{-\phi}}{\NE},0,0,0) $ and $u_\mu= (- e^\phi \NE ,0,0,0)$. 

By requiring stationarity of (\ref{FRWaction}) under small variations of fields, we can proceed to derive the equations of motion. Varying with respect to $s,\phi,\pi_{s},\pi_{\phi},\tilde{N}$ give Hamilton's equations and the Friedmann equation/Hamiltonian constraint:
\begin{widetext}
\begin{align}
 \frac{d\pi_s}{dt} &= 
-  \frac{9K_F e^{3s} }{\NE}  \tilde{h}^2  +  3\kappa \NE e^s
- 3\NE e^{3s}  \LambdaE
+ 3 e^{3s} \left( \frac{\epsilonk}{\NE} y^2 +  \frac{2}{3\tilde{a}_{0}\NE^2} |y|^3 - \NE m_\phi^2 \phi^2  \right) 
\nn
\\
&
+ 8\kgal \frac{e^{3s} y^3 }{\NE^3  } \tilde{h} \left( \kappa e^{-2s} +  \frac{3\tilde{h}^2}{\NE^2}  \right) 
+ 24\pi G\NE e^{3s}  e^{-2\phi} P \label{k_mouflage_pi_s_diff} 
\\
 \frac{d\pi_\phi}{dt} &= - \NE e^{3s}  \left[ 2  m_\phi^2 \phi +  8\pi G  e^{-2\phi} \left( \rho + 3P \right)  \right]. 
\label{k_mouflage_pi_phi_diff}
\\
\frac{ds}{dt} &= \tilde{h}
\\
\frac{d\phi}{dt} &= y
\\
  \frac{3K_F}{\NE^2}  \tilde{h}^2  +  3\kappa e^{-2s}
&= 8\pi G  e^{-2\phi} \rho +    \LambdaE
+ \frac{y^2}{\NE^2}\left( \epsilonk + \frac{4 |y|}{3\tilde{a}_{0}\NE} \right) 
+  m_\phi^2 \phi^2  
+ 8 \kgal \frac{y^3 \tilde{h}}{\NE^4}   \left( 3 \kappa e^{-2s} +  \frac{5 \tilde{h}^2}{\NE^2}  \right) 
\label{k_mouflage_Friedmann_diagonal_frame}
\end{align}
\end{widetext}
where henceforth $\rho$ and $P$ refer to the total matter density and pressure, including a potential matter frame cosmological constant $\Lambda$. Finally, varying with respect to $\tilde{h}$ and $y$ gives the constraint equations
\begin{align}
 \pi_s &=  e^{3s} \left[ -  6K_{F}\frac{\tilde{h}}{\NE}  + 8\kgal \frac{y^3}{\NE^3} \left( \kappa e^{-2s}
   +  \frac{3\tilde{h}^2}{\NE^2}  \right) \right]
 \label{k_mouflage_pi_s_constraint} 
\\
 \pi_\phi &=  e^{3s} \left[  2\epsilonk\frac{y}{\NE}  +  \frac{2}{\tilde{a}_{0}} \frac{|y| y }{\NE^{2}} 
+ 24\kgal \frac{y^2 \tilde{h} }{\NE^3  } \left( \frac{\kappa}{b^2} +  \frac{\tilde{h}^2}{\NE^2}  \right) 
\right] \label{k_mouflage_pi_phi_constraint}
\end{align}
For later use, we introduce the following variables:
\begin{align}
H &\equiv \frac{1}{Na}\frac{da}{dt},\quad \tilde{H} \equiv \frac{1}{\tilde{N}b}\frac{db}{dt},\quad
 \tilde{Y} \equiv \frac{1}{\tilde{N}}\frac{d\phi}{dt}
\end{align}
The quantity $H$ is the rate of change of $\ln a$ with respect to the matter frame proper time $\tau$ where $d\tau=  Ndt$; this is the matter frame Hubble parameter. The quantity $\tilde{H}$ is the Einstein frame Hubble parameter, expressing the rate of change of $\ln b$ with respect to Einstein frame proper time $\tilde{\tau}$ where $d\tilde{\tau}=\tilde{N}dt$ and $\tilde{Y}$ is the rate of change of $\phi$ with respect to $\tilde{\tau}$. Additionally, we find the following relation between the two Hubble parameters:
\begin{equation}
 H =  e^{-\phi}  \left( \HE  - \tilde{Y} \right) 
\label{TeVeS_Hubble_relation}
\end{equation}
Hence $\HE^2   = e^{2\phi} H^2 + 2  e^{\phi}  H \tilde{Y}  + \tilde{Y}^2$ 
so that (\ref{k_mouflage_Friedmann_diagonal_frame}) leads to the matter frame Friedmann equation 
\begin{widetext}
	\begin{align}
	3K_F e^{2\phi} \left[  H^2 + 2  e^{-\phi}  H \tilde{Y}  + e^{-2\phi} \tilde{Y}^2 \right]  
		& = 8\pi G  e^{-2\phi} \rho -  3\kappa e^{-2s}+    \LambdaE
	+ \tilde{Y}^2\left( \epsilonk + \frac{4}{3\tilde{a}_{0}} |\tilde{Y}| \right) 
	+  m_\phi^2 \phi^2  
	\nonumber
	\\
		&+ 8 \kgal   \tilde{Y}^3  (e^\phi H + \tilde{Y} )\left( 3 \kappa e^{-2s} +  5 e^{2\phi} H^2 +  10  e^{\phi}  H \tilde{Y}  +  5 \tilde{Y}^2   \right) 
	\label{k_mouflage_Friedmann_matter_frame}
	\end{align}
\end{widetext}
In finding solutions to the equations of motion, for simplicity we will from now on set $\tilde{\Lambda}=0$ and $\kappa=0$ 
and allow the matter frame cosmological constant $\Lambda$ to be non-zero. 
Additionally we set $m_{\phi}=0$ as was done in \cite{BabichevDeffayetEsposito-Farese2011}.

\subsection{Energy Densities}
\label{eqofstate}
We now look a bit more closely at notions of the energy density of the scalar field $\phi$. This will later aid comparison with the dark sector of the $\Lambda\mathrm{CDM}$ model. It's helpful to write the matter frame Friedmann equation (\ref{k_mouflage_Friedmann_matter_frame})  in a more familiar form:
\begin{equation}
3H^{2} = \frac{8\pi G}{K_{F}}\left(\rho_{r}+\rho_{dust}+\rho_{\Lambda}+\rho_{\phi}\right) 
\label{Normal_Friedmann_equation}
\end{equation}
where
\begin{align}
\rho_{\phi} &\equiv (\rho_{r}+\rho_{dust}+\rho_{\Lambda})(e^{-4\phi}-1)\nn \\
& +\frac{e^{-2\phi}}{8\pi G}  \tilde{Y}^{2} \left(\epsilonk +\frac{4}{3\tilde{a}_{0}}|\tilde{Y}| + 40 \kgal \tilde{Y}\tilde{H}^3\right)
 \nn \\
  &
 +\frac{3K_{F}e^{-2\phi}}{8\pi G}\left(\tilde{Y}^{2}-2\tilde{H}\tilde{Y}\right)
\end{align}
We use the subscript $r$ to denote a radiation component and $\rho_{dust}\equiv \rho_{b}+\rho_{c}$ where $b$ denotes the baryonic contribution and $c$ denotes the contribution due to non-baryonic dust (i.e. a cold dark matter component should one exist).
In writing (\ref{k_mouflage_Friedmann_matter_frame}) in the form (\ref{Normal_Friedmann_equation}) we have entirely separated contributions to $H^{2}$ that depend on $\phi$ and those who do not. 
Furthermore we can develop a notion of fractional energy density of $\Omega_{I}$ of a cosmological component by writing (\ref{Normal_Friedmann_equation}) as
\begin{equation}
	1 = \sum_{I}\Omega_{I}
\qquad
	{\rm where} \quad \Omega_{I} \equiv \frac{8\pi G}{3K_{F}H^{2}}\rho_{I}
\end{equation}
where the index  $I$ is over the various types of matter as well as  $\phi$.
Following earlier TeVeS literature \cite{BourliotEtAl2006}, the Einstein frame Friedmann equation may be written as
\begin{align}
3\tilde{H}^{2} &= \frac{8\pi G}{K_{F}} e^{-2\phi}(\tilde{\varrho}_{r}+\tilde{\varrho}_{dust}+\tilde{\varrho}_{\Lambda}+\tilde{\varrho}_{\phi}) \label{the_Ein_Fried}
\end{align}
where $\tilde{\varrho}_{r}=\rho_{r}$ etc. for matter species and 
\begin{align}
	\tilde{\varrho}_{\phi} &\equiv \frac{e^{2\phi}}{8\pi G} \tilde{Y}^2 \left( \epsilonk +  \frac{4}{3\tilde{a}_{0}} |\tilde{Y}| 
+ 40 \kgal   \tilde{Y} \HE^3 \right) 
\end{align}
This equation may be written as
\begin{equation}
	1=  \sum_{I}\tilde{\omega}_{I}
\quad
        {\rm where} \quad 
	\tilde{\omega}_{I} \equiv  \frac{8\pi G e^{-2\phi}}{3K_{F}\tilde{H}^{2}}\tilde{\varrho}_{I}
\end{equation}
We can rewrite (\ref{the_Ein_Fried})  using \eqref{TeVeS_Hubble_relation} to yield
\begin{align}
3H^{2} &=  8\pi G_{{\rm eff}} \left(\tilde{\varrho}_{r}+\tilde{\varrho}_{dust}+\tilde{\varrho}_{\Lambda}+\tilde{\varrho}_{\phi}\right)  
\label{the_matter_Fried}
\end{align}
where
\begin{equation}
G_{{\rm eff}} =  \frac{ G e^{-4\phi}}{K_{F}\left(1+\frac{d\phi}{d\mathrm{ln}a}\right)^2}
\end{equation}
This equation may be written as 
\begin{equation}
1 = \sum_{I}\omega_{I}
\qquad
        {\rm where} \quad
\omega_{I} \equiv  \frac{8\pi G_{{\rm eff}}}{3H^{2}}\tilde{\varrho}_{I}
\end{equation}
By inspection, 
\begin{equation}
\omega_{I} = \tilde{\omega}_{I}
\end{equation}
The measures of energy density due to the scalar field $\rho_{\phi}$ and $\tilde{\varrho}_{\phi}$ may differ considerably. The presence of the Galileon term implies that the signs of these quantities are not positive-definite and $\rho_{\phi}$ is not positive-definite even in its absence.

\subsection{Numerical Solution Strategy}
Aside from a number of specific situations we will look at, it will be necessary to numerically integrate the equations of motion- this will involve the evolution of the set $\{s,\phi,\pi_{s},\pi_{\phi},\tilde{h},y\}$, i.e.  a total of six variables.
We also have the three constraints \eqref{k_mouflage_Friedmann_diagonal_frame},
\eqref{k_mouflage_pi_s_constraint} and \eqref{k_mouflage_pi_phi_constraint}, hence,
there are three independent variables to be evolved. That is, the space of initial data is three-dimensional.
Finally, we have the four evolution equations \eqref{k_mouflage_HE_diff}, \eqref{k_mouflage_Z_diff}, 
\eqref{k_mouflage_pi_s_diff} and \eqref{k_mouflage_pi_phi_diff} 
for $s,\phi,\pi_{s},\pi_{\phi}$ respectively, but of course not all three are independent.  Additionally, a specific form of $\tilde{N}$ must be chosen.
For integration, first we choose an initial time $t_0$ and set initial conditions for all variables as follows:
\begin{enumerate}
	\item Freely specify values $\{s(t_{0}),\phi_{i}\equiv \phi(t_{0}),y_{i}\equiv y(t_{0})\}$ as initial conditions and choose a form of the function $\tilde{N}$.
	\item Use the constraint \eqref{k_mouflage_Friedmann_diagonal_frame} to find $\tilde{h}(t_{0})$. Of the three roots to this cubic polynomial in $\tilde{h}$, there is one positive root which reduces to the result that would appear in the absence of the Galileon term when we take the limit $\kgal \rightarrow 0$ and we choose this root. If, for example, the contribution of the Galileon term to the right hand side of (\ref{k_mouflage_Friedmann_diagonal_frame}) is small and positive, there will be a positive root $\tilde{h}_{2}\gg \tilde{h}_{1}$ representing another solution to the field equations but it will deviate significantly from $\Lambda\mathrm{CDM}$ and will be neglected. 
		
	\item Use the constraint \eqref{k_mouflage_pi_s_constraint} to determine $\pi_{s}(t_{0})$.
	\item Use the constraint \eqref{k_mouflage_pi_phi_constraint} to determine $\pi_{\phi}(t_{0})$.
\end{enumerate}
At this point, all initial conditions are consistently set.  We proceed to integrate the four equations
\eqref{k_mouflage_HE_diff}, \eqref{k_mouflage_Z_diff},
\eqref{k_mouflage_pi_s_diff} and \eqref{k_mouflage_pi_phi_diff} as follows. For each time $t_n$, repeat the following steps:
\begin{enumerate}
	\item Consider a nearby moment $t_{n+1}> t_{n}$. If $t_{n+1}<t_{\mathrm{end}}$ then evolve to a nearby moment $t=t_{n+1}$ to determine  $\{s(t_{n+1}),\phi(t_{n+1}),\pi_{s}(t_{n+1}),\pi_{\phi}(t_{n+1})\}$ using
	\eqref{k_mouflage_HE_diff}, \eqref{k_mouflage_Z_diff},
	\eqref{k_mouflage_pi_s_diff} and \eqref{k_mouflage_pi_phi_diff}.
	\item Use the constraints \eqref{k_mouflage_pi_s_constraint} and  \eqref{k_mouflage_pi_phi_constraint} to determine $\tilde{h}(t_{n+1})$ and $y(t_{n+1})$. This involves
	solving a pair of coupled equations cubic in $\tilde{h}$ and $y$. Care must be taken that a different branch of $\{\tilde{h},y\}$ solutions is not found here (much like how it was appropriate to find $\tilde{h}_{1}$ rather than $\tilde{h}_{2}$ as a root in the initial value equation for $\tilde{h}$).
	
	\item Check that the constraint \eqref{k_mouflage_Friedmann_diagonal_frame} is satisfied to reasonable numerical precision. This should guarantee that the 
	previous step has avoided skipping onto a different branch of solutions for $\{\tilde{h},y\}$.
\end{enumerate}

\subsection{Regimes and Approximate Solutions}
\label{regimes}

We now consider the behavior of the scalar field and metric in some specific situations. For ease of calculation we'll use a spacetime gauge where $\tilde{N}=1$ and so we identify $t=\tilde{\tau}$ where $\tilde{\tau}$ measures Einstein frame cosmic  proper time. The evolution equation for $\pi_{\phi}$ (\ref{k_mouflage_pi_phi_diff}) can be integrated to yield:
\begin{align}
\pi_{\phi} &=  \pi_{\phi(0)}+\int_{\tilde{\tau}_{0}}^{\tilde{\tau}}Jd\tilde{\tau} 
\\
J &\equiv  -8\pi G b^{3}e^{-2\phi}(\rho+3P)
\end{align}
where again $\rho$ and $P$ refer to the total matter densities and pressures and $\pi_{\phi(0)}$ is a constant.
From the $\pi_{\phi}$ constraint \eqref{k_mouflage_pi_phi_constraint}
and defining $\signY=\sign[\tilde{Y}]$,  we then have:
\begin{equation}
{\cal A}(\tilde{\tau})\tilde{Y}^{2}+\tilde{Y} +{\cal C}(\tilde{\tau}) = 0 
\label{peq} 
\end{equation}
 where
\begin{align}
{\cal A}(\tilde{\tau}) 
&\equiv \frac{1}{\epsilonk}\left(\frac{\signY}{\tilde{a}_{0}}+12\kgal\tilde{H}^{3}\right) 
\label{nonlinscale}
\\
{\cal C}(\tilde{\tau}) 
&\equiv 
- \frac{1}{2\epsilonk b^{3}} \left(\pi_{\phi(0)} + \int_{\tilde{\tau}_{0}}^{\tilde{\tau}}Jd\tilde{\tau} \right)
\label{linscale_C}
\end{align}
Solutions to (\ref{peq}) are the cosmological analog of the solution (\ref{dphidr}), in other words,
we are dealing with a temporal Vainshtein mechanism. 

We now identify two main regimes of evolution.  When 
\begin{equation}
\tilde{Y} \sim -{\cal C} \label{linear} \equiv \tilde{Y}_{{\rm lin}}
\end{equation}
we shall refer to this as the  \emph{Canonical Regime} as here the canonical kinetic term is dominating the evolution of the field $\phi$. When 
\begin{equation}
\tilde{Y}^{2} \sim -\frac{\cal C}{\cal A} \equiv \tilde{Y}^{2}_{{\rm nonlin}}  
\label{nonlinear}
\end{equation}
we will refer to this as the \emph{Nonlinear Regime}. The transition between the linear and nonlinear regimes occurs when $\tilde{Y}_{lin}\sim \tilde{Y}_{nonlin}$ i.e. when $|{\cal C}{\cal A}| \sim 1$ and the Nonlinear Regime corresponds to $|{\cal C}{\cal A}| \gg 1$.

 Within the Nonlinear Regime limit, there exist two further distinct sub-regimes, depending on the dominant terms inside ${\cal A}$. The first of these 
 is when
\begin{equation} 
\left|\frac{\signY}{\tilde{a}_{0}}\right| \gg \left|12\kgal \tilde{H}^{3}\right|
\end{equation}
In this limit the MOND term contribution to ${\cal A}$ dominates and hence we will refer this to the \emph{MOND Regime}.

 Alternatively, when
\begin{equation}
\left|\frac{\signY}{\tilde{a}_{0}}\right| \ll \left|12\kgal\tilde{H}^{3}\right|
\end{equation}
then ${\cal A}(\tilde{\tau})$ is dominated by the Galileon term and so we refer to this as the \emph{Galileon Regime}. 

The g-TeVeS parameter restrictions of Section \ref{Section_Parameterspace} enforce that ${\cal A}(\tilde{\tau})$ is positive-definite during the Galileon era. Therefore, if these restrictions are adopted then the solution (\ref{nonlinear}) only possesses real solutions if ${\cal C}$ is negative. By inspection $J$ is negative-definite for matter sources with equation of state $w\geq -1$ and so the function ${\cal C}$ may only be non-positive with a suitably chosen $\pi_{\phi(0)}$. 

Consider two universes with identical ${\cal C}(t)$ and with parameters enabling similar evolution of the metric but with one with a sufficiently large ${\cal A}(t)$ to push it into the nonlinear regime.
The ratio of $\tilde{Y}$ in the universe where it is in the nonlinear regime to that of a universe where it is in the linear regime is:
\begin{equation}
\frac{\tilde{Y}_{{\rm nonlin}}}{\tilde{Y}_{{\rm lin}}} \sim \sqrt{\frac{-1}{{\cal A}_{{\rm nonlin}}{\cal C}}}
 \label{pratio}
\end{equation}
As $|{\cal A}_{{\rm nonlin}}{\cal C}| \gg 1$, then the ratio (\ref{pratio}) is below unity so that $\tilde{Y}_{{\rm nonlin}}$ is suppressed 
compared to $\tilde{Y}_{lin}$. This is a realization of a temporal Vainshtein mechanism. In particular as $b\rightarrow 0$, then $\HE \rightarrow\infty$ so that $\tilde{Y} \rightarrow 0$ and the theory reduces to GR in the early universe.

We now discuss some approximate solutions to the theory; these will aid the interpretation of numerical results.

\subsubsection{Canonical Regime with matter fluid with constant $w_{mat}$}
\label{approxcanon}

First we consider the limit where the scalar field is inside the Canonical Regime and there is a single matter source with constant equation of state $w_{mat}\equiv P_{mat}/\rho_{mat}$. Consider the following ansatz:
\begin{equation}
\phi = n\ln b + \phi_{0} 
\label{canon_phi_ansatz}
\end{equation}
Then we have $\tilde{Y} = n \tilde{H}$ and it is found that the evolution equations for $\pi_{\phi}$ and $\pi_{s}$ subject to the Hamiltonian constraint are consistent if 
\begin{equation}
n = \frac{(1+3w_{mat})}{(w_{mat}-1)}\frac{K_{F}}{\epsilon_{c}}
\end{equation}
which yields an Einstein frame Friedmann equation:
\begin{align}
\left(1-\frac{\epsilon_{c}n^{2}}{3K_{F}}\right)\tilde{H}^{2}= \frac{8\pi G e^{(1+3w_{mat})\phi}\rho_{(mat)0}}{3K_{F} b^{3(1+w_{mat})}}
\end{align}
It follows then that
\begin{equation}
\frac{\tilde{\varrho}_{\phi}}{\tilde{\varrho}_{\phi}+\rho_{mat}} 
  = \tilde{\omega}_{\phi}= \frac{(1+3w_{mat})^{2}K_{F}}{3(w_{mat}-1)^{2}\epsilonk} 
\label{canontrack}
\end{equation}
The interpretation of this is that the scalar field tracks the matter component so that $\tilde{\varrho}_{\phi}/\rho_{mat}$ is a constant.
This situation is identical to ordinary TeVeS~\cite{SkordisEtAl2005,BourliotEtAl2006}.

Recall now that $H=e^{-\phi}(\tilde{H}-\tilde{Y}) = e^{-\phi}(1-n)\tilde{H}$ and so 
$\tilde{H}^{2}= e^{2\phi}H^{2}/(1-n)^{2}$, and also $b=e^{\phi}a$, and so 
\begin{equation}
3H^{2}= 8\pi G_{eff}\frac{\rho_{(mat)0}}{a^{3(1+w_{mat})}}
\end{equation}
where we have defined an effective Newton's constant $G_{eff}$:

\begin{align}
G_{eff} &\equiv  \frac{e^{-4\phi}G}{K_{F}}\frac{(1-n)^{2}}{\left(1-\frac{\epsilon_{c}n^{2}}{3K_{F}}\right)} \label{canon_geff}
\end{align}
Note that $G_{eff}$ is not positive-definite and will generally have a time dependence via $\phi$. In the dust-dominated  regime ($w_{mat}=0$) and cosmological constant-dominated regime ($w_{mat}=-1$), the effective Newton's constant in the Canonical regime is negative for $0 < \epsilon_{c}/K_{F} < 1/3$ and this should be excluded.

We see from (\ref{canon_geff}) that in the Canonical Regime, the ratio $\epsilonk/K_{F}$ controls two important effects. The first is a constant rescaling of the effective Newton's constant in the matter frame Friedmann equation (i.e. the explicit dependence of $G_{eff}$ on $n$), the second is a time-dependent rescaling of the effective Newton's constant via the factor $e^{-4\phi}$. From (\ref{canon_phi_ansatz}) we see that $d\phi/d\ln b$ is proportional to $n$ and so a larger value of $K_{F}/\epsilonk$ will lead to a more dramatic growth in $\phi$.

\subsubsection{General Regime with negligible gravitation of scalar field kinetic terms}
Again we consider a dominant gravitating matter source with constant $w_{mat}$. Now assume that there is a situation where $\tilde{Y}$ has a negligible impact on the Einstein-frame Friedmann equation \eqref{k_mouflage_Friedmann_diagonal_frame}
 and evolution equation \eqref{k_mouflage_pi_s_diff} for $s$ i.e. that

\begin{align}
\frac{d\pi_{s}}{d\tilde{\tau}} &\simeq  24\pi G b^{3}e^{-2\phi} (w_{mat} - 1 ) \rho_{w} 
\label{christmas1} 
\\
3K_{F}\tilde{H}^{2} &\simeq 8\pi G e^{-2\phi}\rho_{w}
\end{align}
where $\rho_{w}\equiv \rho_{(w)0}/a^{3(1+w_{mat})}$. Moreover \eqref{k_mouflage_pi_phi_diff} gives
\begin{align}
\frac{d\pi_{\phi}}{d\tilde{\tau}}&= -8\pi G (1+3w_{mat}) b^{3}e^{-2\phi}\rho_{w} 
\label{christmas2}
\end{align}
The above equations suggest a linear relation between $\pi_\phi$ and $\pi_s$ so that  we may take the ansatz $\pi_{\phi} =\beta \pi_{s} + C_\pi$
-- where $C_\pi$ is a constant-- and comparing (\ref{christmas1}) and (\ref{christmas2}) we have that
\begin{equation}
\beta = \frac{(1+3w_{mat})}{3(1-w_{mat})}
\end{equation}
while from \eqref{k_mouflage_pi_s_constraint} and \eqref{k_mouflage_pi_phi_constraint} and having in mind 
 that $|K_{F}\tilde{H}| \gg 4|\kgal \tilde{Y}^3 \HE^2|$ 
\begin{align}
\frac{\pi_\phi}{b^3} =2\epsilonk\tilde{Y}  +  \frac{2}{\tilde{a}_{0}} |\tilde{Y}| \tilde{Y} + 24\kgal \tilde{Y}^2 \HE^3 
\nonumber
\\
 \simeq 
 \frac{2(1+3w_{mat})}{(w_{mat}-1)} \HE K_F  +  \frac{C_\pi}{b^3} 
\label{approximate_solution}
\end{align}
%
 Given this, in the Canonical Regime for $C_\pi=0$ we have $\tilde{Y}\appropto \tilde{H}$ which is consistent with the results of 
Section \ref{approxcanon}.  

In the non-linear regime we find the following behaviour:
\paragraph{MOND regime}
In the MOND Regime we find two distinct behaviors which depend on the size of the constant $C_\pi$. If $3|\beta K_F \HE| \gg |C_\pi| b^{-3}$ 
 and $\HE>0$ we have $\tilde{Y}\simeq -\sign(\beta) \sqrt{3 \tilde{a}_{0} |\beta| K_F} \HE^{1/2}$. In the radiation era this corresponds to
$\tilde{Y} \sim 1/a$ and $\phi \sim C_1 + C_2 a$ while in the matter era it corresponds to
$\tilde{Y} \sim a^{-3/4}$ and $\phi \sim C_1 + C_2 a^{3/4}$~\footnote{The constants $C_1$ and $C_2$ are of course not the same in all
of the considered cases.}.
 In both cases, the evolution of $\phi$ is negligible with respect 
to the Friedman equation where $\phi$ can be taken to be approximately constant.

If $3|\beta K_F \HE| \ll |C_\pi| b^{-3}$ then $\tilde{Y} \sim a^{-3/2}$ irrespective of the equation of state of matter. However,
 the evolution of $\phi$ in the radiation era is  $\phi \sim C_1 + C_2\sqrt{a}$ while in the matter era
 $\phi \sim C_1 +   C_2 \ln a$. The later, is akin to the tracker solutions found in section  \ref{approxcanon}.

It is easy to check that both types of behaviors are consistent and may also be derived using \eqref{nonlinscale}
 and \eqref{linscale_C} into \eqref{nonlinear} under the {\it a posteriori} justified assumption $\phi \sim C_1$.

\paragraph{Galileon regime}
 In the Galileon Regime we find once again two distinct behaviors which depend on the size of the constant $C_\pi$.
If for $3|\beta K_F \HE| \gg |C_\pi| b^{-3}$ we have $\tilde{Y}^{2}\tilde{H}^{2} \simeq -\beta K_F/(4\kgal)$.
Clearly, this case can only be satisfied for $\kgal < 0$. Furthermore, during the radiation era
$\tilde{Y} \sim a^{2}$ and $\phi \sim C_1 + C_2 a^{4}$ while in the matter era
$\tilde{Y} \sim a^{3/2}$ and $\phi \sim C_1 + C_2 a^{3}$.

 If   $3|\beta K_F \HE| \ll |C_\pi| b^{-3}$ we have that $\tilde{Y} \sim a^{3/2}$ 
and $\phi \sim C_1 + C_2 a^{7/2}$ while in the matter era 
$\tilde{Y} \sim a^{3/4}$ and $\phi \sim C_1 + C_2 a^{9/4}$.

In all cases, the constants $C_2$ multiplying the time-dependent part of $\phi$ are tiny and are in addition multiplied by tiny values 
of the scale factor, so that $\phi$ may in fact be taken to be constant during the galileon era.

\subsection{Comparison To $\Lambda\mathrm{CDM}$}
\label{lcdm_comparison}
A given set of parameters and initial conditions of the model will produce an expansion history $H(a)$. The behavior of this function will reveal  the extent to which the model can imitate the effect of dark matter on the cosmological background expansion. A simple measure of the extent to which this agrees with the $\Lambda\mathrm{CDM}$ `concordance model' expansion history $H_{\Lambda\mathrm{CDM}}(a)$ is via the quantity
\begin{align}
{\cal S} &\equiv 
   \frac{\int^{a_{n}}_{a_{i}}\left(\frac{H-H_{\Lambda\mathrm{CDM}}}{H_{\Lambda\mathrm{CDM}}}\right)^{2}d\ln a}{\int^{a_{n}}_{a_{i}}d\ln a}
 \label{s_quantity}
\end{align}
where $a_{i}$ is the physical scale factor at the beginning of integration of the equations of motion (taken to be deep in the radiation era),  $a_{n}$ is its value today and $H_{\Lambda CDM}(a)$ is determined by General Relativity's Friedmann equation
\begin{align}
3H_{\Lambda CDM}^{2} &= 8\pi G_{N}\left(\rho_{c}+\rho_{b}+\rho_{r}+\rho_{\Lambda}\right)
\end{align}
where spatial curvature has been assumed to be zero. For the fiducial $\Lambda\mathrm{CDM}$ model to which we compare to,
 we use the following cosmological parameters:
\begin{align}
\Omega_{c}(a_{n}) &= 0.264
\nn\\
\Omega_{b}(a_{n}) &=0.0492\nn\\
\Omega_{r}(a_{n}) &= 9.25\times 10^{-5}\nn\\
\Omega_{\Lambda}(a_{n}) &= 1-\Omega_{c}-\Omega_{b}-\Omega_{r}
\end{align}
The $\Lambda\mathrm{CDM}$ Hubble parameter $H_{0}$ today is taken to be $67 \mathrm{km} \mathrm{s}^{-1}\mathrm{Mpc}^{-1}$. We use units where this value is equal to unity 
and hence, in particular, we have $H_{0}/a_{0}\simeq 5.44$. We take the scale factor today $a_{n}$ to also be equal to unity 
and in looking at g-TeVeS models we take $\rho_{b}$ and $\rho_{r}$ in general to be assumed known and identical to their corresponding 
values in the $\Lambda\mathrm{CDM}$ case.

Exploration of how ${\cal S}$ varies across the parameter space of the k-mouflage model will help us understand what features of the model are responsible for producing a similar or disimilar cosmology to that of $\Lambda \mathrm{CDM}$. 
The parameters of the g-TeVeS model cosmological background evolution equations are as follows:

\begin{enumerate}
	\item $\epsilonk$: the canonical kinetic term coefficient
	\item $\tilde{a}_{0}$: the MOND kinetic term coefficient
	\item $\kgal$: the Galileon kinetic term coefficient
	\item $K_{F}$: Einstein-aether field constant.
	\item $\phi_{i}(a_{i})$: the initial value of the scalar field at the start of integration 
	\item $y_{i}(a_{i})$: the coordinate time derivative $d\phi/dt$ at the start of integration.
\end{enumerate}
Of the matter content, we allow $\Lambda$ and- in some cases- $\rho_{c}$ to vary and for simplicity we assume that spatial curvature $\kappa$ is negligible.
Consider the matter frame Friedmann equation at some early time deep in the radiation era:
\begin{equation}
3H^{2} = \frac{8\pi G e^{-4\phi_{i}}}{K_{F}}\frac{\rho_{r0}}{a^{4}}+ \dots
\end{equation}
where the dots denote assumed subdominant contributions due to other species of matter and time derivatives of $\phi$. If the Friedmann equation takes this form across a span of time where $\tilde{Y} \ll H$ deep into the radiation era then the expansion of the universe will be approximately that of General Relativity with a value of Newton's constant $G_{C} =  G e^{-4\phi_{i}}/K_{F}$. 
This implies though that $G_{C}$ will not necessarily be the same as the locally measured value of Newton's constant $G_{N}$ which itself is assumed in \cite{BabichevDeffayetEsposito-Farese2011} to be equal to $G$. We will continue this assumption in the optimization procedure.
Constraints on the primordial helium abundance from Big Bang Nucleosynthesis (BBN) \cite{CarrollLim2004} constrain the difference between these two quantities to be in the following interval:
\begin{equation}
\frac{7}{8} < \left(\frac{G_{C}}{G_{N}}\right)^{2} < \frac{9}{8}
\end{equation}
This should be regarded as a conservative constraint; it is likely that using restrictions from BBN in conjunction with data from the cosmic microwave background (CMB) will be able to place much more stringent restrictions on $G_{C}/G_{N}$ \cite{AvilezSkordis2014}. If the above assumptions hold then the above bounds constrain the deviation of $e^{-4\phi_{i}}/K_{F}$ from unity and this is implemented in the procedure to try and find minima of the quantity (\ref{s_quantity}). Instead of varying the pair $\{\phi_{i},K_{F}\}$ we equivalently 
vary $\{\phi_{i},G_{C}\}$.

The final quantity that we will consider is the (assumed dust-like) dark matter density $\rho_{c}$ (or equivalently the constant ratio $\rho_{c}/\rho_{b}$). For a modified gravity alternative to dark matter, it may be hoped that the new gravitational degrees of freedom of the theory may producde a dark matter effect by themselves; failing that, it is instructive to see how much dark matter may be needed in addition (not necessarily the amount present in $\Lambda\mathrm{CDM}$). 

We consider the following four models:

\begin{itemize}
	\item Model A: Minimization of ${\cal S}$ given the restrictions on the g-TeVeS theory's parameters described in Section \ref{Section_Parameterspace} and the restriction that $\rho_{c}=0$.
	\item Model B: Minimization of ${\cal S}$ given the restrictions on the g-TeVeS theory's parameters
	described in \ref{Section_Parameterspace} and an unrestricted $\rho_{c}$
	\item Model C: Minimization of ${\cal S}$ without any restrictions on the g-TeVeS theory's parameters and the restriction that $\rho_{c}=0$
	\item Model D: Minimization of ${\cal S}$ without any restrictions on the g-TeVeS
	theory's parameters and an unrestricted $\rho_{c}$.
\end{itemize}
To minimize ${\cal S}$ we employ a Markov Chain Monte-Carlo (MCMC) procedure where random initial 
 values for $V=\{\epsilonk,\tilde{a}_{0},\kgal,G_{C},\Lambda,\rho_{c}/\rho_{b},\phi_{i},y_{i}\}$ are chosen subject to the 
restrictions placed upon the particular model then new values are generated and adopted if they lead to a decrease in ${\cal S}$.

\begin{figure}
\center
\epsfig{file=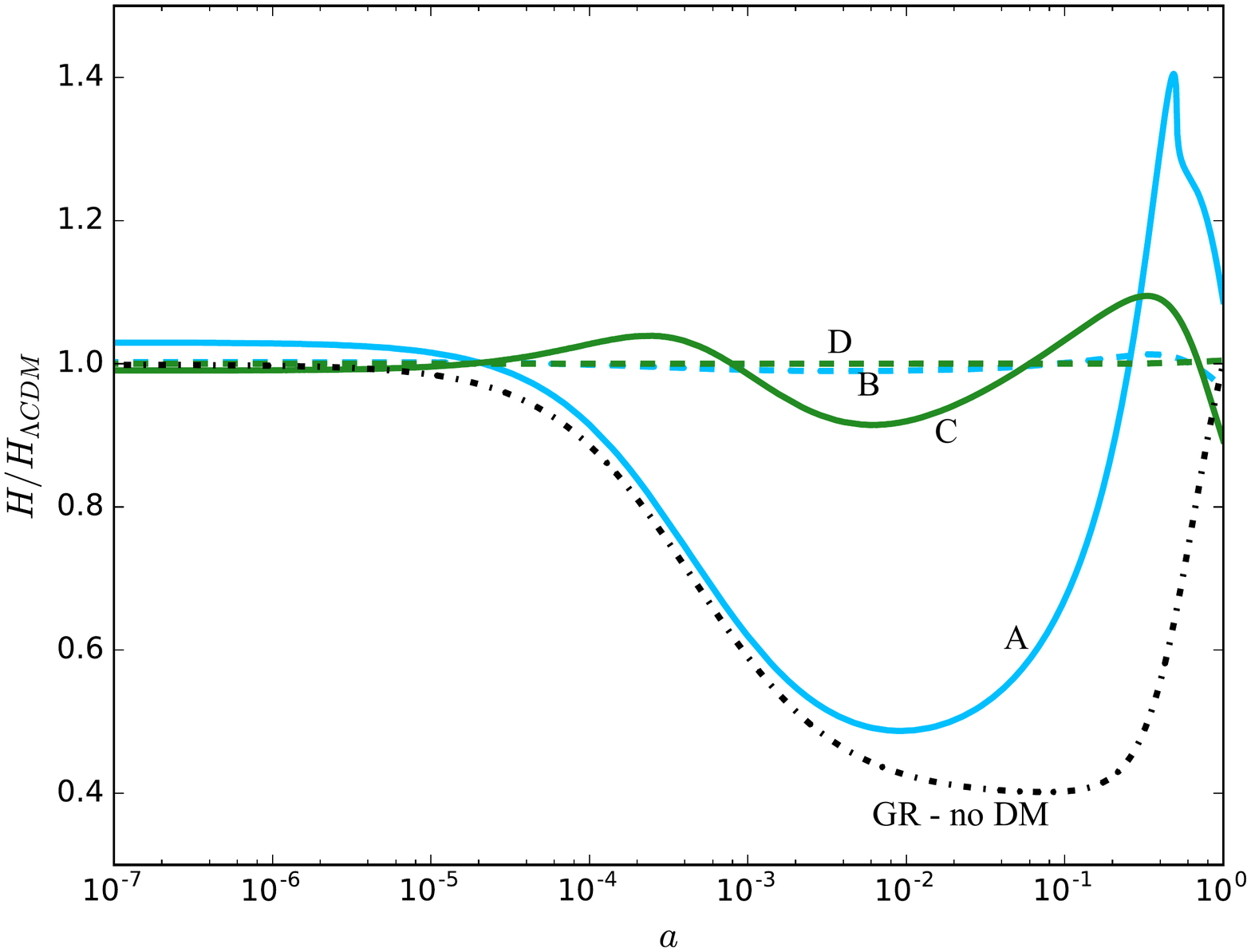,width=9.cm}
\caption{The ratio of physical Hubble parameter $H(a)$ to the $\Lambda\mathrm{CDM}$ Hubble parameter $H_{\Lambda\mathrm{CDM}}$ 
for best fit models using MOND phenomenology restrictions (blue curves) and without those restrictions (green curves). Solid curves 
represent fits obtained without additional dark matter $\rho_{c}$ whilst dashed curves represent fits allowing for a non-zero $\rho_{c}$.
The later are visually almost indistinguishable from the $\Lambda\mathrm{CDM}$ prediction. The black dotted curve
is the corresponding plot for a universe described by GR with no dark matter and $\Omega_{\Lambda}=1-\Omega_{r}-\Omega_{b}$.}
\label{hubbleratios}
\end{figure}

\renewcommand\arraystretch{1.4}
\begin{table*}
		\centering
	\caption{Best-fit values for various models, quoted to two significant figures: Model A: Restricted parameters with $\rho_{c}=0$; Model B: Restricted parameters with $\rho_{c} \neq 0$; Model C: Unrestricted parameters with $\rho_{c} =0$; Model D: Unrestricted parameters with $\rho_{c} \neq 0$.}
	\label{table_of_models}
	\begin{tabular}{|c|ccccccccc|}
			\hline
		 & ${\cal S}$ & $\epsilonk$ & $1/\tilde{a}_{0}$ & $8\kgal$ & $\Lambda$ & $\rho_{c}/\rho_{b}$ &$\phi_{i}$ & $y_{i}$ & $G_{C}/G-1$ \\ \hline
$A$	& $7.0 \times 10^{-2}$           &    $4.7  \times 10^{-2}$        & $5.4$          & $1.6\times 10^{-24}$        &              $2.3$     &           $0.0$     &$6.0\times 10^{-2}$ &  $1.2\times 10^{-8}$ & $6.0\times 10^{-2}$  \\ \hline
	$B$	& $3.9 \times 10^{-5}$           &      $1.6 \times 10^{-13}$      &  $5.4$         &  $1.0 \times 10^{-25}$       &          $1.6$         & $5.2$     &$1.4$&    $5.6 \times 10^{-9}$   & $4.8\times 10^{-3}$        \\ \hline
$C$	& $1.7 \times 10^{-3}$           & $2.7 \times 10^{1}$            &  $1.7 \times 10^{-4}$         & $-7.2 \times 10^{-39}$            &      $1.3\times 10^{-1}$             &   $0.0$  &$8.9 \times 10^{-3}$&  $-3.1 \times 10^{-5}$   & $-1.9\times 10^{-2}$        \\ \hline
$D$	&   $9.9 \times 10^{-7}$         &        $2.3 \times 10^{13}$    &     $8.0 \times 10^{-1}$      & $-6.7 \times 10^{-35}$           &     $2.1$              &          $5.4$  &$2.4\times 10^{-5}$& $-1.1 \times 10^{-8}$  & $-2.1\times 10^{-4}$     \\ \hline
	\end{tabular}
\end{table*}
The best-fit parameters for each model are contained in Table \ref{table_of_models} and for each of the best-fits given, the ratio of their predicted $H(a)$ vs physical scale factor $a$ is plotted in Figure \ref{hubbleratios}. We additionally show the corresponding plot for the 
General Relativistic case without Dark Matter and with $\Omega_{\Lambda}= 1-\Omega_{b}-\Omega_{r}$ and this corresponds to a value ${\cal S} =0.11$. 
We stress that we make no comparison with data and that these ``best-fits'' are with respect to closeness to our fiducial $\Lambda\mathrm{CDM}$ model.
We now discuss the features of the best-fit for each model:

\paragraph{Model A}
The best-fit value of ${\cal S}$ for this model is $0.07$, and so in the sense of (\ref{s_quantity})  the Hubble parameter is somewhat closer to $\Lambda\mathrm{CDM}$ than the General Relativistic model without dark matter. However, as can be seen from Figure \ref{hubbleratios} , the best-fit model has a Hubble parameter substantially smaller than that of $\Lambda\mathrm{CDM}$ for much of the cosmological matter era, meaning that the scalar field $\phi$ is unable to provide an additional dust-like dark matter component during this era. 

It may be verified that for the entirety of the evolution of $\phi$, the field $\tilde{Y}$ is in the Nonlinear Regime with $|\tilde{Y}| \gg |{\cal C}{\cal A}|$ (see Section \ref{regimes}), being in the Galileon Regime until around $a\sim 10^{-5}$ before transitioning to the MOND regime; the feature where the $H/H_{\Lambda CDM}$ momentarily is greater than unity shortly before $a\sim 1$ corresponds to $\tilde{Y}$ changing sign but quickly returning to the MOND regime.
It was found in Section \ref{approxcanon} that if the theory is in the Canonical Regime that the effective Newton's constant $G_{eff}$ is negative for $0<\epsilonk/K_{F} < 1/3$; given that $\epsilonk$ is restricted to be less than $0.1$ for Model A, it is possible that this tends to an avoidance of the Canonical Regime. We will see in the case of Model C that the theory being in the Canonical Regime for much of the matter era is vital to producing a dark matter effect.

\paragraph{Model B}
This model retains the parameter restrictions of Model A but allows for a non-zero $\rho_{c}$. The best-fit value of ${\cal S}$ for this model is $3.9\times 10^{-5}$, considerably closer to $\Lambda\mathrm{CDM}$ than that of Model A; inspection of Figure \ref{hubbleratios} shows the corresponding physical Hubble parameter $H(a)$ to be extremely similar to that to the $\Lambda\mathrm{CDM}$ Hubble parameter. From Table \ref{table_of_models} we see that a value of $\rho_{c} = 5.2\rho_{b}$ is preferred- a contribution of dark matter comparable to that in $\Lambda CDM$. 

It may be verified that the behavior of $|\tilde{Y}|$ for this fit is roughly similar to that of Model A and hence the theory remains in the Nonlinear Regime throughout cosmological evolution but interestingly Model B is the one model of the four considered here which appears to favour a value of $\phi_{i}$ of order unity.

\begin{figure}[h]
\epsfig{file=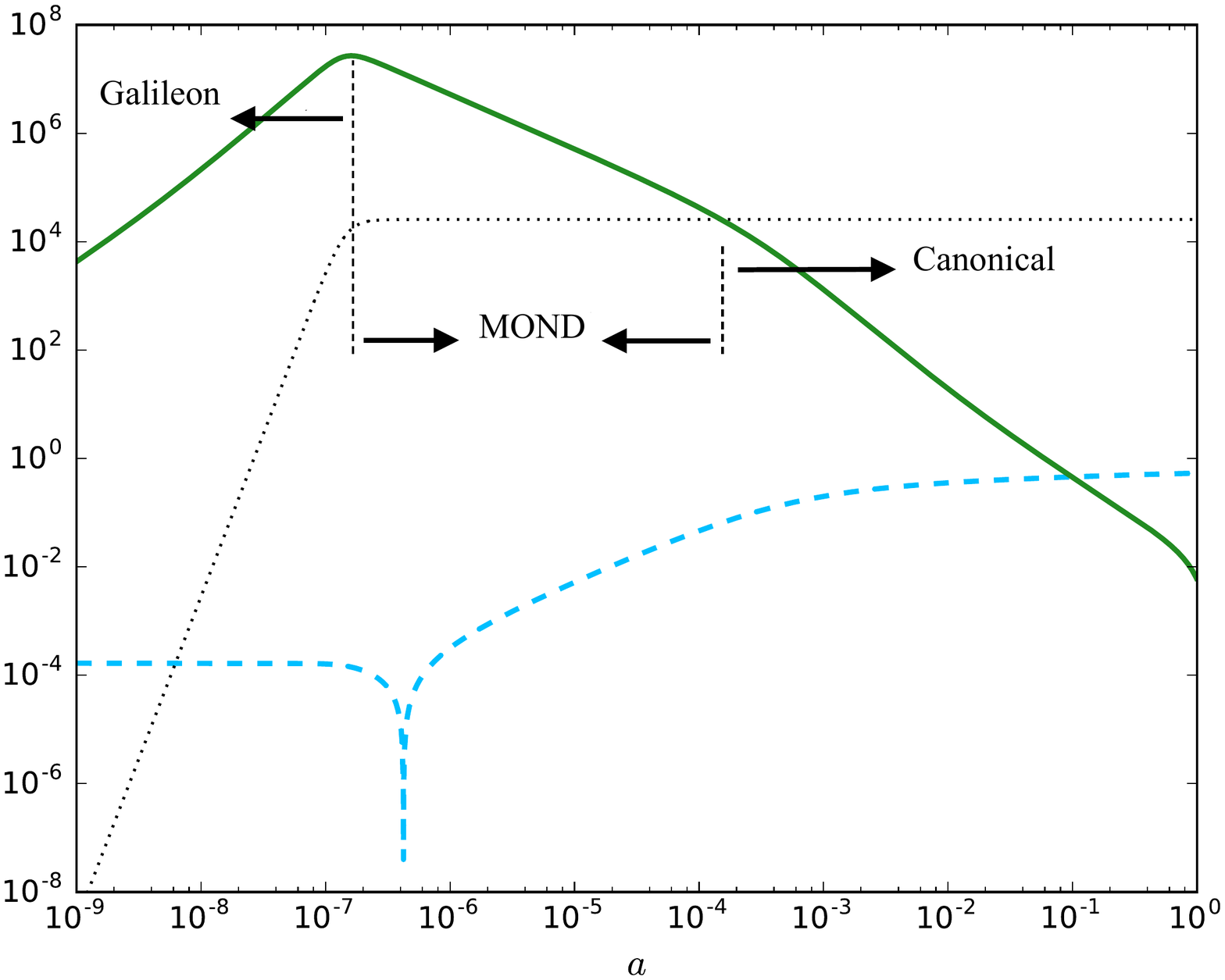,width=9cm}
	\caption{Evolution of $-\tilde{Y}$ (solid green), the quantity $1/|{\cal A}|$ (dotted black), and $-\phi$ (dashed blue) 
for Model C best-fit constant parameters. Sensitivity of $\tilde{Y}$ to the dominant term in $1/|{\cal A}|$ visible 
whilst the field $\phi$ can be seen to deviate significantly from its initial value over the universe's evolution.}
	\label{nonlinearscale}
\end{figure}

\paragraph{Model C}
For this model the astrophysical restrictions are not used on the parameter space and there is
 no additional dark matter. A value of ${\cal S} = 1.7\times 10^{-3}$ is found and hence the closeness to the $\Lambda\mathrm{CDM}$ model compared to Model A is rather greater.
 For this model, the evolution of $\tilde{Y}$ and $\phi$ as a function of physical scale factor $a$ are plotted in Figure \ref{nonlinearscale} as well as the quantity $1/|{\cal A}|$ of (\ref{nonlinscale}) which tracks the dominant source of nonlinearity; this quantity increases during the Galileon Regime and is approximately constant during the MOND Regime. We see then from Figure \ref{nonlinearscale} that the theory is initially in the Galileon Regime and during this time $\tilde{Y}$ grows as $1/\tilde{H} \sim a^{2}$ (using 
 the approximate proportionality of scale factors $b$ and $a$  due to the near-constancy of $\phi$ during this era); this evolution is in accordance with the results of (\ref{approximate_solution}), suggesting that its assumptions such as the linear relation between scalar field momentum $\pi_{\phi}$ and gravitational momentum $\pi_{s}$ are good ones. If the solution (\ref{approximate_solution}) is an accurate approximation then we see that in the radiation era the scalar field in the Galileon Regime acts like an effective cosmological constant in the Einstein frame Friedmann equation (\ref{k_mouflage_Friedmann_diagonal_frame}). At later times during the radiation era, the scalar field transitions to the MOND Regime and Figure \ref{nonlinearscale} shows $\tilde{Y}$ decreasing as $\tilde{H}^{1/2}\sim a^{-1}$ here, again consistent with the results of (\ref{approximate_solution}) and whilst these approximations hold then during this regime, terms in $\tilde{Y}$ produce an additional, dust-like contribution to (\ref{k_mouflage_Friedmann_diagonal_frame}).
 
 As the dominant non-scalar field contribution to the background evolution becomes the baryonic dust, $\rho_{\phi}$ will no longer vary as $a^{-3}$ during the MOND Regime within the approximation (\ref{approximate_solution}) and the optimization procedure prefers parameters that produce a transition to the Canonical Regime at roughly the time when baryons begin to dominate over radiation; this leads to a preferred value of $\tilde{a}_{0}\sim 5.9\times 10^{3}$ and this value is substantially different to the MOND acceleration scale $a_{0} \simeq 5.44$. However, we emphasize that the relation between $\tilde{a}_{0}$ and $a_{0}$ may be such that they need not be numerically close.

From the solutions of Section \ref{approxcanon}, we see that in the limit of constant $\phi$, the scalar field produces an 
effective rescaling of Newton's constant and thus acts like an additional gravitating source whose density varies as $a^{-3}$. 
Crucially though, $\phi$ is found to vary substantially during the era where the density of baryons is greater than that of radiation 
(see Figure \ref{nonlinearscale}). The effective dark matter-like contribution to $H$ in the matter era is then produced by a 
time-dependent rescaling of the effective Newton's constant.
From (\ref{canon_phi_ansatz}) we see that the rate of growth of $\phi$ is determined by the ratio $K_{F}/\epsilonk$; this ratio must 
take the right value so that the rescaling of Newton's constant due to $\phi$ is not too big or too small. The best-fit favors 
a value $\epsilonk/K_{F}\sim 27$. Additionally, the best-fit involves a negative value of $\kgal$; we note however that similar but 
slightly higher values of ${\cal S}$ can be reached with positive $\kgal$ and so it is not clear how much cosmology favors one 
sign of $\kgal$ over another. Bear in mind, however, that the astrophysical Vainshtein mechanism operates only for $\kgal>0$ and so negative $\kgal$ are of dubious
importance.

\paragraph{Model D}
Finally we consider a model which involves no restrictions on the g-TeVeS parameters and allows for a dark matter component. An excellent 
fit is achieved, with ${\cal S}= 9.9\times 10^{-7}$ and, as with Model B, it can be seen from Figure \ref{hubbleratios} that the physical 
Hubble parameter $H(a)$ is extremely close to $H_{\Lambda\mathrm{CDM}}(a)$. The preferred values for $\rho_{c}$ and $\Omega_{\Lambda}$ are 
close to the values chosen for the $\Lambda\mathrm{CDM}$ cosmology used. This indicates that the  modification of gravity due to the 
fields $\phi$ and $A_{\mu}$ has been suppressed.

It can be verified that for this best-fit, the field $\tilde{Y}$ enters the Canonical  Regime deep in the radiation era and never 
spends time in the MOND Regime; this is in part due to the large value of $\epsilonk\sim 2.3\times 10^{13}$ and this value also 
leads to a far smaller gravitational contribution of $\tilde{Y}$, as we expect from the results of Section \ref{approxcanon}.

\subsubsection{The initial data for $\tilde{Y}$ and the existence of pathological solutions}
\begin{figure}[h]
	\epsfig{file=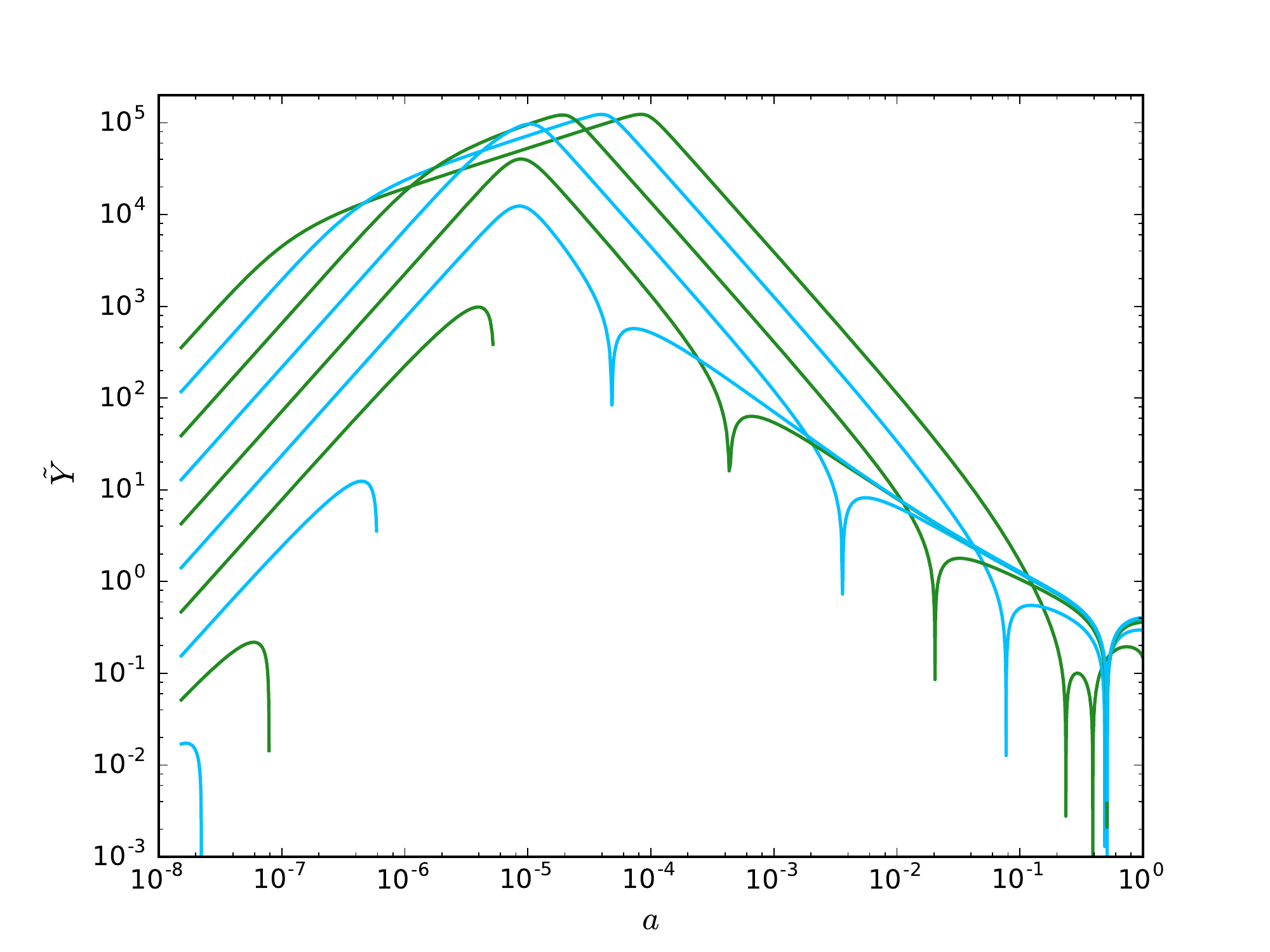,width=8.7cm}
	\caption{Evolution of $|\tilde{Y}|$ for various initial, positive values in the case where the constants correspond to best-fit parameters for Model A. Lower values of $\tilde{Y}_{i}$ (all of which are positive numbers) display evolution that is thwarted at some $a<1$; this is observed to coincide in each case with  $\tilde{Y}^{2}$ attempting to pass through zero and happens only during Galileon domination. All other curves persist but belong to universes that ultimately  (shortly after $a=1$) recollapse.}
	\label{cliffhanger}
\end{figure}

It is found that a wide variety of $\{\epsilonk,\tilde{a}_{0},\kgal,G_{C},\Lambda,\rho_{c}/\rho_{b},\phi_{i},y_{i}\}$ yield universes that do not reach the defined present cosmological moment $a=1$. This is illustrated in Figure (\ref{cliffhanger}) which shows the evolution of $|\tilde{Y}|$ for varying initial values of $|\tilde{Y}|$ and for parameters that obey the restrictions of constants of the Model A best-fit. Each of the curves that cease continuing past a particular scale factor $a < 1$ represent a cosmos that cannot evolve past that scale factor; this is not due to the expansion rates $H$ or $\tilde{H}$ getting stuck at zero but rather because the equations of motion no longer possess real solutions at the following moments. Curves whose evolution is thwarted are those where $|\tilde{Y}|$- having initially been positive- reaches zero whilst the Galileon term dominates over the MOND term; curves which persist to evolve after reaching $0$ are those which reach $0$ when the MOND term dominates.
Roughly speaking, in the limit of Galileon domination, the evolution equation for $\tilde{Y}$ may be written in the form:

\begin{equation}
\frac{d\tilde{Y}^{2}}{dt} = \xi(t) 
\label{galileon_z_equation}
\end{equation}
hence $\tilde{Y}^{2}(t) = \tilde{Y}_{i}^{2} + 2\int_{t_{i}}^{t} \xi dt$. If the integral contribution is negative and decreasing as $t$ increases then at some time $t_{0}$, $\tilde{Y}^{2}$ will reach zero and then evolution will cease as $\tilde{Y}^{2}$ cannot continue on to take negative values.
If, instead, the MOND term dominates then the evolution equation for $\tilde{Y}$ takes the form

\begin{equation}
|\tilde{Y}|\frac{d\tilde{Y}}{dt} = \zeta(t)
\end{equation}
This does not encounter the same problem as that with (\ref{galileon_z_equation}). Upon reaching $0$ following an initial, positive $\tilde{Y}_{i}$,  $\tilde{Y}$ can pass to a negative value, for which the solution takes the form $-\tilde{Y}^{2}(t)=2\int_{t_{0}}^{t} \zeta dt$, where $t_{0}$ is the moment that $\tilde{Y}$ reaches $0$, and so a continually negative $\zeta(t)$ does not cause evolution to stop.

Even for curves which persist, it can be verified that of them in Figure \ref{cliffhanger} correspond to universes that eventually begin collapsing. The values of $\{\epsilonk,\tilde{a}_{0},\kgal,G_{C},\Lambda,\rho_{c}/\rho_{b},\phi_{i},z_{i}\}$ are such for these curves that the collapse of the universe for these universes begins after the present moment $a=1$.

\begin{figure}
	\epsfig{file=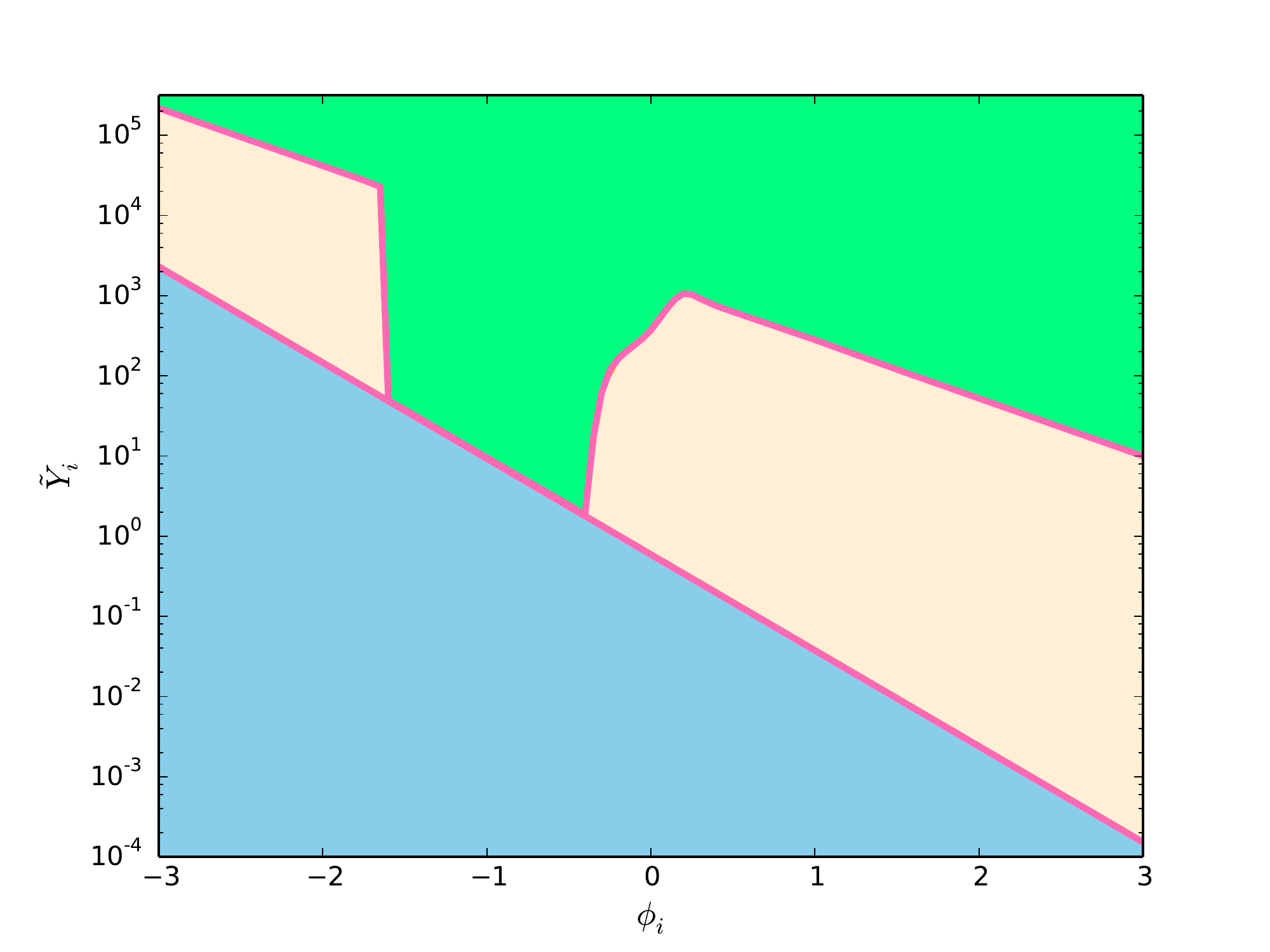,width=8.7cm}
	\centering
	\caption{Plot of classification of universe evolution for differing initial data $\{\phi_{i},\tilde{Y}_{i}\}$. The green (top) region denotes universes that collapse before the universe reaches $a=1$; the blue (lower) region denotes 
	universes which experience thwarted evolution; the orange (middle) region denotes universes that persist to $a=1$.}
	\label{regionofstability}
\end{figure}

\begin{figure}[h]
	\epsfig{file=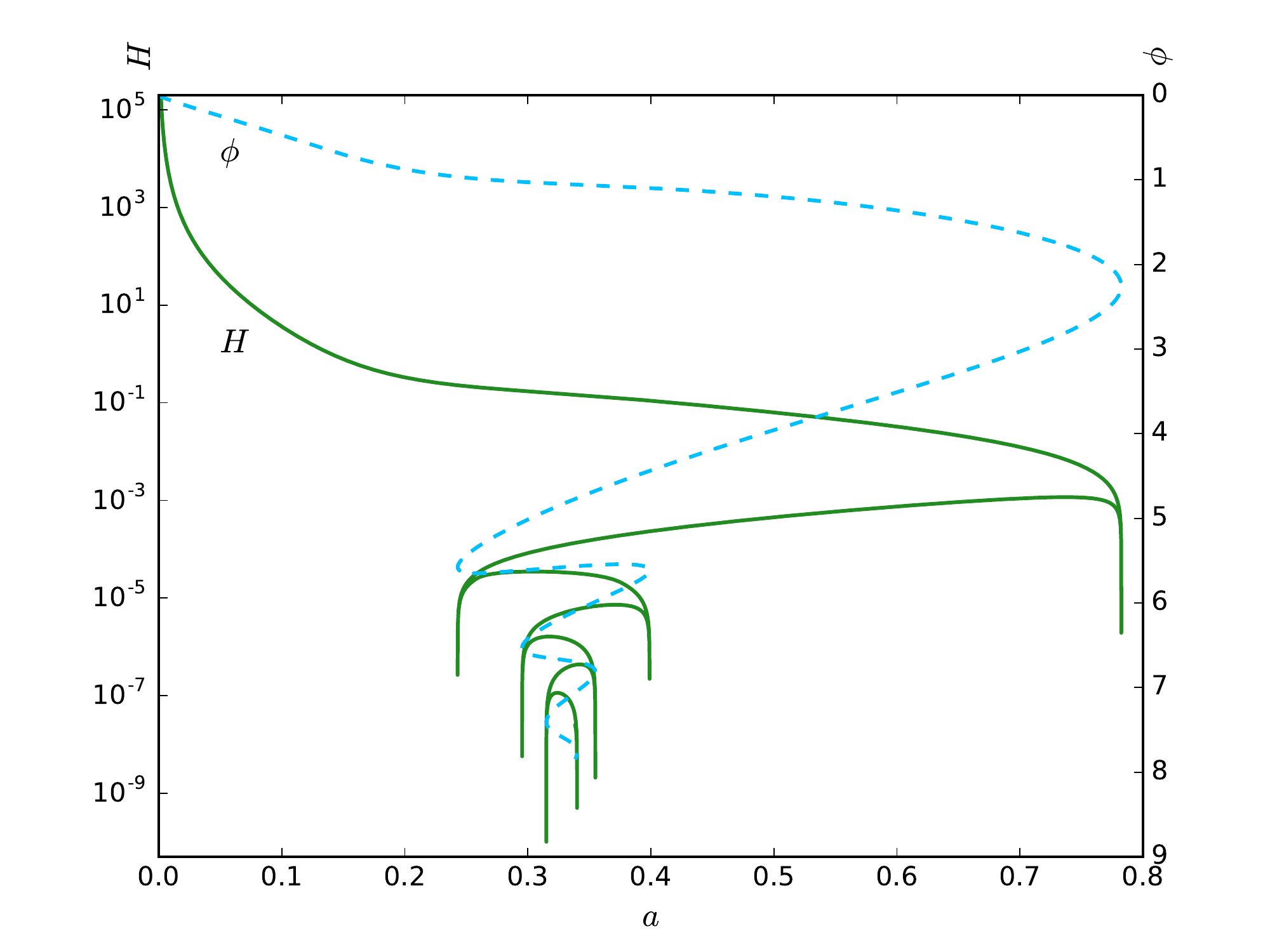,width=8.7cm}
	\caption{Solution with Model A best-fit constant parameters but with initial data on $\tilde{Y}$ adjusted so as to give a universe which 
collapses before the present day. Unusual damped-oscillatory behaviour can be seen in the scale factor $a$ continuing up to the end of 
numerical integration. }
	\label{waterfall}
\end{figure}

As can further be seen from Figure \ref{cliffhanger}, many curves at early times follow a solution $|\tilde{Y}|\sim b^{3/2}$ which is consistent with the approximate solution (\ref{approximate_solution}) assuming $\phi \sim \mathrm{cst.}$ (hence $a\propto b$) and $\tilde{H} \sim 1/b^{2} \sim 1/a^{2}$ i.e. that the universe is radiation dominated. For the larger values of $|\tilde{Y}_{i}|$  we see a deviation from this solution before $|\tilde{Y}|$ begins decreasing. The decreasing of $|\tilde{Y}|$ marks the onset of the MOND regime and the deviation from $|\tilde{Y}|\sim a^{3/2}$ marks a new solution where $|\tilde{Y}|\sim a^{1/2}$ and is having a significant impact on the background evolution. It is found that higher yet initial values of $|\tilde{Y}|$ to collapse of the universe before $a=1$. These different behaviors are shown in Figure (\ref{regionofstability}) for a wide range of initial data $\{\phi_{i},\tilde{Y}_{i}\}$ for the Model A best-fit constant parameters. Initial data in the lower (blue) region of the plot all lead to thwarted evolution. Initial data in the upper (green) region correspond to universes where the universe expands for a period but collapses before $a=1$. The upper region is constructed from two, independent possibilities: the region above the (interrupted) upper straight line in Figure (\ref{regionofstability}) denotes parameter space where no real, positive solutions to the Einstein frame Friedmann equation (\ref{k_mouflage_Friedmann_diagonal_frame}) exist; the additional region which splits the middle region where evolution persists to the present scale factor involves solutions where the universe evolves for some time with positive $\tilde{h}$ but where ultimately the physical scale factor $H$ passes through zero to a negative value at a moment earlier than the present. The accompanying, unusual, evolution of $H$ for one such universe is shown explicitly in Figure \ref{waterfall}.

Negative initial $\tilde{Y}_{i}$ for Model A parameters are found to generally lead to similar behavior 
but with a substantially narrower range of parameters leading to universes which persist. For less 
restricted parameter ranges- particularly those with $\kgal<0$- these restrictions on the allowable $\tilde{Y}_{i}$ don't 
necessarily still apply; the reason for this is likely that when $\kgal<0$ then in the Galileon Regime there exist 
solutions of (\ref{approximate_solution} )for $\tilde{Y}^{2}$ even when $\xi=0$ - which is not the case for $\kgal>0$.

The presence of thwarted evolution and (too early) collapse of universes represents a challenge to the optimization procedure as many universes become simply unsuitable for comparison with $\Lambda CDM$.


\subsection{Comparison to TeVeS}
\label{Section_TeVeS}
Having seen FRW solutions of the k-mouflage Galileon model, we now look at how results compare to those of the original TeVeS theory. 
This will help show what comparative features of g-TeVeS and TeVeS are responsible for dark matter-type effects in cosmology. As was 
discussed in Section \ref{Section_Introduction}, the two theories differ in their form of the scalar field action $S_{\phi}$; an 
equivalent formulation of TeVeS has the following form of scalar field action:

\begin{align}
	S_{\phi}[\phi,\mu] &=  -\frac{1}{16\pi G} \int d^{4}x\sqrt{-\tilde{g}}
 \left[\mu\tilde{g}^{\mu\nu}\tilde{\nabla}_{\mu}\phi\tilde{\nabla}_{\nu}\phi \right. 
\nn 
\\
	& \left.+W(\mu)\right]
\end{align}
where the auxiliary field $\mu$ has been has been introduced.
In FRW symmetry we have:

\begin{equation}
	S_{\phi } =  \frac{1}{16\pi G}\int dt e^{3s}\tilde{N} \left[\mu\frac{y^{2}}{\tilde{N}^{2}}-W(\mu)\right]
\end{equation}
From this we have contributions to the $s$,$\tilde{N}$,$y$ equations of motion and a new equation of motion-obtained by varying with respect to $\mu$:

\begin{equation}
	\frac{y^{2}}{\tilde{N}^{2}} - \frac{dW(\mu)}{d\mu} = 0
\end{equation}
It may then be possible to obtain $\mu(y)$ from this equation. 
In \cite{Bekenstein2004}, Bekenstein proposed a specific choice of $W$ which yielded MOND-like phenomenology on astrophysical scales. This form is: 

\begin{align}
	W(\mu) &= \xi_B\left[\muh\left(4+2\muh-4\muh^2+\muh^3\right)+4\ln|\muh-1|)\right]
\end{align}
where $\muh \equiv \mu/\mu_{0}$, where $\mu_{0}$ and $\xi_{B}$ are constant parameters~\footnote{The constant
$\xi_B$ is related to Bekenstein's constant $\ell_B$ and $\mu_0$ 
via $\xi_B =  \frac{\mu_0^3}{64\pi \ell_B^2}$, according to the conventions of
~\cite{BourliotEtAl2006}.}. The FRW cosmological regime spans from $\muh=2$ to $+\infty$. 
Applying the optimization procedure to this model- and again assuming $G=G_{N}$~\footnote{Strictly speaking $G\ne G_N$ in TeVeS, but 
rather $G_N/G = 2/\mu_0 + 2/(2- c_1+c_4)$. However, as we are interested in comparing to g-TeVeS where the exact relation 
is at present unknown, we keep using $G=G_N$.  In any case, the exact relation between the two constants will only become important 
when placing constraints using data, which is not what we do in the present work.}- we find that ${\cal S}$ 
is minimized for the following parameter values:

\begin{align*}
	{\cal S}&= 2.4\times 10^{-3}, \mu_{0} = 24, \xi_{B} =2.6\times 10^{7}, \nn 
\\
& 	y_{i} = -3.3\times 10^{-11}, 
 \Lambda = 0.17, 
\nn 
\\
 &\phi_{i} = -3.1\times 10^{-4},  \left(\frac{G_{C}}{G}-1\right) = 3.2\times 10^{-2}
\end{align*}
where all dimensionful parameters are expressed in units where $H_0=1$.

\begin{figure}
	\epsfig{file=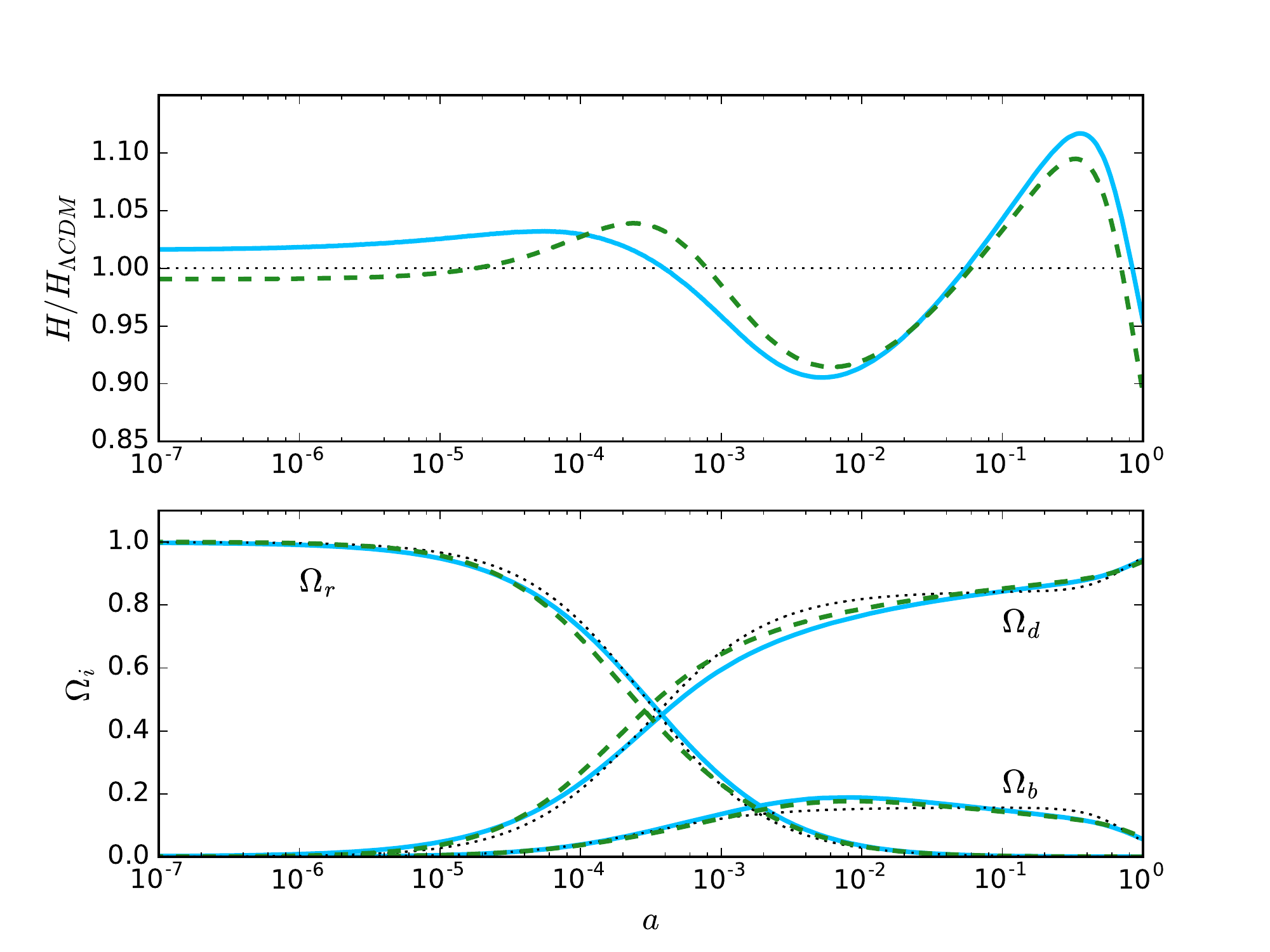,width=8.7cm}
	\caption{Plots of $H/H_{\Lambda\mathrm{CDM}}$ and $\Omega_{I}$ for best-fit TeVeS model (solid blue), 
best-fit unrestricted g-TeVeS model with $\rho_{c}=0$ (dashed green), and $\Lambda\mathrm{CDM}$ (dotted black). $\Omega_{b}$ is 
the baryon contribution; $\Omega_{r}$ the radiation contribution; $\Omega_{d}$ the contribution due to the 
dark sector: dark matter, cosmological constant, and scalar field.}
	\label{tevesdensity}
\end{figure}

\begin{figure}
	\epsfig{file=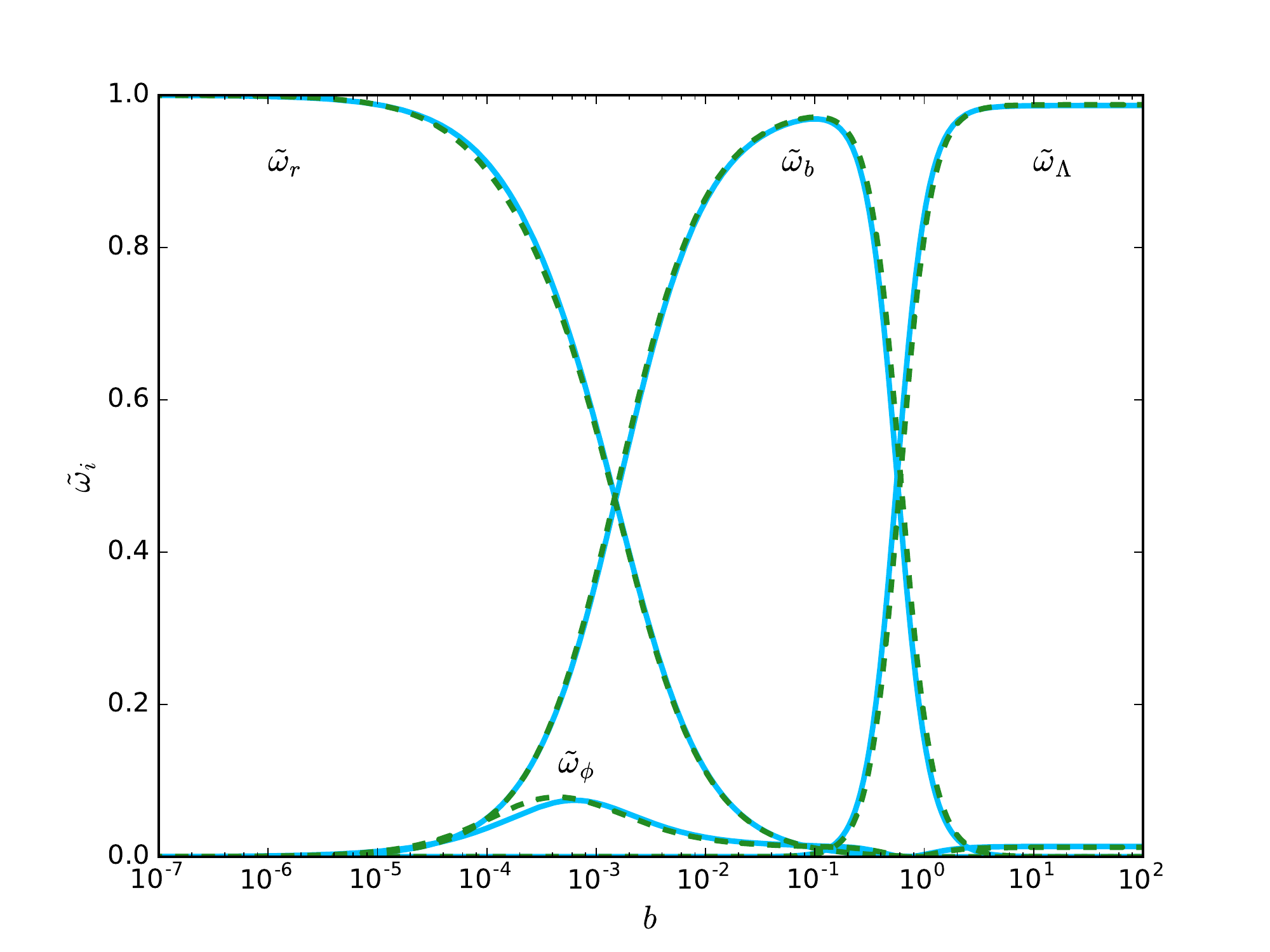,width=8.7cm}
	\caption{Plots of Einstein frame quantities $\tilde{\omega}_I= \omega_I$ for best-fit TeVeS model (blue solid lines) 
and best-fit unrestricted g-TeVeS model (green dashed lines). Both sets of curves are very similar. Tracking behavior of the 
scalar field $\phi$ is evident in the matter and $\Lambda$ dominated era.}
	\label{teveseindensity}
\end{figure}

This is a comparatively good fit to the Model C best-fit unrestricted g-TeVeS model and it can be seen from Figures \ref{tevesdensity} 
and \ref{teveseindensity} that each fit produces similar expansion histories and $\Omega_{I}$, $\tilde{\omega}_{I}$ quantities. 
It is apparent from Figure \ref{teveseindensity} that TeVeS reduces to a model dominated by a canonical kinetic term at around 
between $b=10^{-3}$ and $10^{-2}$ (when the baryons begin to dominate the Einstein frame expansion) due to the tracking of the 
dominant matter component in agreement with equation (\ref{canontrack}).. It may be shown that in the 
Canonical Regime of TeVeS, the number $\mu_0/K_F$ corresponds to $\epsilonk/K_F$; for the TeVeS best-fit and the Model C g-TeVeS
best-fit (see Section \ref{lcdm_comparison}), these quantities take similar values (approximately $25$ and $27$ respectively).
Furthermore, the value of $\phi$ at $a=1$ is similar in both cases: $-0.527$ (TeVeS) and $-0.532$ (g-TeVeS).

\subsection{Conclusions and Discussion}
In this paper we have examined the solutions to the g-TeVeS theory in FRW symmetry. The theory has been constructed to replicate the success of MOND which is a modification to Newtonian gravity and as such involves only visibile matter and the Newtonian gravitational field. The relativistic theory, however, involves the introduction of new degrees of freedom in the gravitational sector and these can not just complicate the relation between constants appearing in actions and observed quantities but they can also modify MOND itself.
A manifestation of the former is that the numbers $G$ and $a_{0}$ associated with MOND may depend on the local value for the scalar field $\phi$. We have seen that if the theory is to have a dark matter-type effect in the background cosmology then this will likely involve significant growth of $\phi$ over cosmological history. Therefore it may be inaccurate to assume that $|\phi|\ll 1$ on astrophysical scales and this would endanger the assumptions made in \cite{BabichevDeffayetEsposito-Farese2011} that the bare Newton's constant $G$ and acceleration scale $\tilde{a}_{0}$ are to be identified with the measured Newton's constant and MOND acceleration scale.

It is found that the theory can give a reasonable match to the $\Lambda\mathrm{CDM}$ expansion history for parameters $\{\epsilonk,\kgal,\tilde{a}_{0}\}$ that differ orders of magnitude from the suggested restrictions on them given in \cite{BabichevDeffayetEsposito-Farese2011} however it remains an open question whether these values (accompanying the substantial deviation of $\phi$ from $0$) can lead to MOND-like phenomenology.

We can isolate three effects that seem relevant to the theory resembling a cold dark matter component. The first is that when the gravitational effect of $\phi$ is sub-dominant in the radiation era, if the theory is in the MOND Regime, then the scalar field's kinetic terms approximate an additional dust component; this does not persist into the (baryonic) matter era and so the MOND Regime must be exited and the Canonical Regime reached. In this regime, the growth of $\phi$ is substantial and this can lead to an increase in Newton's constant. If the growth is of the right amount, then this increase can resemble- by making matter heavier roughly by a factor of the $\Lambda\mathrm{CDM}$ ratio $(\rho_{b}+\rho_{c})/\rho_{b}$- an additional cold dark matter component. If approximation to $\Lambda\mathrm{CDM}$ pushes the theory towards being in the MOND regime during much of the radiation era tends to suppress $\kgal$ relative to $\tilde{a}_{0}$; indeed similarly good fits for Model C were possible if $\kgal = 0$. The cosmological moment of transition between MOND Regime and Canonical Regime is further dictated by $\epsilonk$ and $\tilde{a}_{0}$ so their values must be tuned to find the right moment. The amount of rescaling of Newton's constant chiefly depends on $\epsilonk/K_{F}$ and so it must be tuned in isolation.

Though it was not possible to find cosmologies similar to that of $\Lambda\mathrm{CDM}$ when parameters were restricted to obey the astrophysical constraints of \cite{BabichevDeffayetEsposito-Farese2011}, one may wonder whether additional terms may be added to the action (\ref{kmouflage_action}) to modify the cosmological behaviour of the theory. We have investigated the effect of allowing for a non-zero $m_{\phi}$ and it appears that this term does not significantly improve the best-fit in the case of the restricted parameters of Model A. A reason for this is perhaps that even in the absence of MOND and Galileon terms in (\ref{kmouflage_action}), the scalar sector of a canonical kinetic term together with a mass term is not equivalent to adding a massive scalar field because of the different gravitational effect of the scalar field in Einstein and matter frames; though the contribution of the scalar field to the Einstein frame Friedmann equation (\ref{k_mouflage_Friedmann_diagonal_frame}) may approximate that of a massive scalar field in some limits, this is not the case for the matter frame Friedmann equation (\ref{k_mouflage_Friedmann_matter_frame}) with `cross' terms such as $H\tilde{Y}$ present. Due to the presence of the aether field $A_{\mu}$, one can build additional terms to add to the gravitational Lagrangian that project out time-like variations of $\phi$; for example, a term $A^{\mu}\tilde{\nabla}_{\mu}\phi A^{\nu}\tilde{\nabla}_{\nu}\phi|A^{\sigma}\tilde{\nabla}_{\sigma}\phi|$
will behave identically to the MOND kinetic term in the background cosmology but conceivably will have only a marginal effect on astrophysical scales. However, it is not clear how to consistently
add or not add such terms to (\ref{kmouflage_action}).

We have additionally checked that similar behavior existed in the original TeVeS theory. For Bekenstein's original choice of free function, the theory possesses a limiting form describing a Canonical Regime and the theory's parameters can be chosen such that a similar amount of rescaling of Newton's constant as in best-fit g-TeVeS models can occur and it is found that parameters of the theory may be chosen that this regime exists for much of the matter era. Prior to the matter era, the theory transitions to a  Nonlinear Regime which differs from the MOND Regime and does not give an additional dust-like contribution during the radiation era. This is perhaps the reason why the g-TeVeS theory gives a slightly closer fit to $\Lambda\mathrm{CDM}$ than TeVeS with Bekenstein's function but the effect is small, as can be seen from the similarity of the results shown in Figures \ref{tevesdensity} and \ref{teveseindensity}.

An important step in further investigating the cosmological consequences of g-TeVeS will be to develop the theory of cosmological perturbations as was done for the TeVeS theory \cite{SkordisEtAl2005,Skordis2005}. This will make it possible to deduce the effect of the new gravitational degrees of freedom on the cosmological microwave background and growth of large scale structure and will likely provide a powerful test of the theory.

\begin{acknowledgments}
	We  thank St{\' e}phane Ili{\' c} and Michael Kopp for many useful discussions. The research leading to these results has received funding from the European Research Council under the European Union's Seventh Framework Programme (FP7/2007-2013) / ERC Grant Agreement n. 617656 ``Theories and Models of the Dark Sector: Dark Matter, Dark Energy and Gravity.''
\end{acknowledgments}

\appendix

\subsection{Quantities in FRW Symmetry}
\label{Appendix_FRW}
Some useful quantities can be calculated:

\begin{align}
\tilde{\Gamma}^{t}_{tt} &= \frac{1}{\tilde{N}}\frac{d\tilde{N}}{dt},  \quad\quad \tilde{\Gamma}^{t}_{ij} = \frac{1}{2\tilde{N}^{2}}\frac{d b^{2}}{dt}
\\
\tilde{\Gamma}^{i}_{tj} &= \frac{1}{b}\frac{db}{dt}\delta^{i}_{\ph{i}j}, \quad\quad \tilde{\Gamma}^{i}_{jk} = \gamma^{i}_{jk}
\end{align}
where $\gamma^{i}_{jk}$ are the Christoffel symbols corresponding to the metric $\gamma_{ij}$ and spatial derivatives acting on it.
From this it follows that:

\begin{align}
\tilde{R}_{tktl} &= -\tilde{N}^{2}\gamma_{kl}\left[\frac{d}{dt}\left(\frac{b}{\tilde{N}^{2}}\frac{db}{dt}\right)
 +\frac{b}{\tilde{N}^{3}}\frac{d\tilde{N}}{dt}\frac{db}{dt}\right.
 \nn 
\\
& 
\left.-\frac{1}{\tilde{N}^{2}}\frac{db}{dt}\frac{db}{dt}\right]
 \\
\tilde{R}_{ijkl} &=  b^{2}\left(\kappa+\frac{1}{\tilde{N}^{2}}\frac{db}{dt}\frac{db}{dt}\right) 
 \left(\gamma_{ik}\gamma_{lj}-\gamma_{il}\gamma_{kj}\right)
\end{align}
For the scalar field ansatz $\phi=\phi(t)$, the only non-vanishing part of $\tilde{\nabla}_{\mu}\phi$ will be $\tilde{\nabla}_{t}\phi=d\phi/dt$. Given our previous results, it follows that:

\begin{align}
\tilde{\nabla}_{t}\tilde{\nabla}_{t}\phi &= \frac{d^{2}\phi}{dt^{2}}
\\
\tilde{\nabla}_{i}\tilde{\nabla}_{j}\phi &= 
 -\tilde{\Gamma}^{t}_{ij}\frac{d\phi}{dt} = -\frac{b}{\tilde{N}^{2}}\frac{db}{dt}\frac{d\phi}{dt}\gamma_{ij}
\end{align}
All other components of $\tilde{\nabla}_{\mu}\tilde{\nabla}_{\nu}\phi$ being zero. Additionally we have that:

\begin{equation}
\tilde{\nabla}_{i}A_{j} = \tilde{N}\tilde{\Gamma}^{t}_{ij} = \frac{1}{2\tilde{N}}\frac{db^{2}}{dt}\gamma_{ij}
\end{equation}
with all other components of $\tilde{\nabla}_{\mu}A_{\nu}$ vanishing.

\bibliographystyle{unsrtnat}
\bibliography{references}

\begin{thebibliography}{50}
\providecommand{\natexlab}[1]{#1}
\providecommand{\url}[1]{\texttt{#1}}
\expandafter\ifx\csname urlstyle\endcsname\relax
  \providecommand{\doi}[1]{doi: #1}\else
  \providecommand{\doi}{doi: \begingroup \urlstyle{rm}\Url}\fi

\bibitem[Milgrom(1983)]{Milgrom1983}
M.~Milgrom.
\newblock {A Modification of the Newtonian dynamics as a possible alternative
  to the hidden mass hypothesis}.
\newblock \emph{Astrophys. J.}, 270:\penalty0 365--370, 1983.
\newblock \doi{10.1086/161130}.

\bibitem[Milgrom(1999)]{Milgrom1998}
Mordehai Milgrom.
\newblock {The modified dynamics as a vacuum effect}.
\newblock \emph{Phys. Lett.}, A253:\penalty0 273--279, 1999.
\newblock \doi{10.1016/S0375-9601(99)00077-8}.

\bibitem[Klinkhamer and Kopp(2011)]{KlinkhamerKopp2011}
F.~R. Klinkhamer and M.~Kopp.
\newblock {Entropic gravity, minimum temperature, and modified Newtonian
  dynamics}.
\newblock \emph{Mod. Phys. Lett.}, A26:\penalty0 2783--2791, 2011.
\newblock \doi{10.1142/S021773231103711X}.

\bibitem[Pazy and Argaman(2012)]{PazyArgaman2011}
E.~Pazy and N.~Argaman.
\newblock {Quantum particle statistics on the holographic screen leads to
  Modified Newtonian Dynamics (MOND)}.
\newblock \emph{Phys. Rev.}, D85:\penalty0 104021, 2012.
\newblock \doi{10.1103/PhysRevD.85.104021}.

\bibitem[Sanders and McGaugh(2002)]{SandersMcGaugh2002}
Robert~H. Sanders and Stacy~S. McGaugh.
\newblock {Modified Newtonian dynamics as an alternative to dark matter}.
\newblock \emph{Ann. Rev. Astron. Astrophys.}, 40:\penalty0 263--317, 2002.
\newblock \doi{10.1146/annurev.astro.40.060401.093923}.

\bibitem[Bekenstein and Magueijo(2006)]{BekensteinMagueijo2006}
Jacob Bekenstein and Joao Magueijo.
\newblock {Mond habitats within the solar system}.
\newblock \emph{Phys. Rev.}, D73:\penalty0 103513, 2006.
\newblock \doi{10.1103/PhysRevD.73.103513}.

\bibitem[Gentile et~al.(2007)Gentile, Famaey, Combes, Kroupa, Zhao, and
  Tiret]{GentileEtAl2007}
G.~Gentile, B.~Famaey, F.~Combes, P.~Kroupa, H.~S. Zhao, and O.~Tiret.
\newblock {Tidal dwarf galaxies as a test of fundamental physics}.
\newblock \emph{Astron. Astrophys.}, 472:\penalty0 L25, 2007.
\newblock \doi{10.1051/0004-6361:20078081}.

\bibitem[Wu et~al.(2009)Wu, Zhao, Wang, Llinares, and Knebe]{WuEtAl2009}
Xufen Wu, HongSheng Zhao, Yougang Wang, Claudio Llinares, and Alexander Knebe.
\newblock {N-body simulations for testing the stability of triaxial galaxies in
  MOND}.
\newblock \emph{Mon. Not. Roy. Astron. Soc.}, 396:\penalty0 109, 2009.
\newblock \doi{10.1111/j.1365-2966.2009.14735.x}.

\bibitem[Dai et~al.(2010)Dai, Matsuo, and Starkman]{DaiStarkman2010}
De-Chang Dai, Reijiro Matsuo, and Glenn Starkman.
\newblock {Limited utility of Birkhoff's theorem in modified Newtonian
  dynamics: Nonzero accelerations inside a shell}.
\newblock \emph{Phys. Rev.}, D81:\penalty0 024041, 2010.
\newblock \doi{10.1103/PhysRevD.81.024041}.

\bibitem[Zhao and Famaey(2010)]{ZhaoFamaey2010}
HongSheng Zhao and Benoit Famaey.
\newblock {Comparing different realizations of modified Newtonian dynamics:
  virial theorem and elliptical shells}.
\newblock \emph{Phys. Rev.}, D81:\penalty0 087304, 2010.
\newblock \doi{10.1103/PhysRevD.81.087304}.

\bibitem[Magueijo and Mozaffari(2012)]{MagueijoMozaffari2011}
Joao Magueijo and Ali Mozaffari.
\newblock {The case for testing MOND using LISA Pathfinder}.
\newblock \emph{Phys. Rev.}, D85:\penalty0 043527, 2012.
\newblock \doi{10.1103/PhysRevD.85.043527}.

\bibitem[Famaey and McGaugh(2012)]{FamaeyMcGaugh2011}
Benoit Famaey and Stacy McGaugh.
\newblock {Modified Newtonian Dynamics (MOND): Observational Phenomenology and
  Relativistic Extensions}.
\newblock \emph{Living Rev. Rel.}, 15:\penalty0 10, 2012.
\newblock \doi{10.12942/lrr-2012-10}.

\bibitem[Hees et~al.(2016)Hees, Famaey, Angus, and Gentile]{Heesetal2015}
A.~Hees, B.~Famaey, G.~W. Angus, and G.~Gentile.
\newblock {Combined Solar System and rotation curve constraints on MOND}.
\newblock \emph{Mon. Not. Roy. Astron. Soc.}, 455\penalty0 (1):\penalty0
  449--461, 2016.
\newblock \doi{10.1093/mnras/stv2330}.

\bibitem[Margalit and Shaviv(2016)]{MargalitShaviv2015}
Ben Margalit and Nir~J. Shaviv.
\newblock {Constraining MOND using the vertical motion of stars in the solar
  neighbourhood}.
\newblock \emph{Mon. Not. Roy. Astron. Soc.}, 456\penalty0 (2):\penalty0
  1163--1171, 2016.
\newblock \doi{10.1093/mnras/stv2721}.

\bibitem[Pereira et~al.(2016)Pereira, Overduin, and
  Poyneer]{PereiraOverduinPoyneer2016}
Jonas~P. Pereira, James~M. Overduin, and Alexander~J. Poyneer.
\newblock {Satellite Test of the Equivalence Principle as a Probe of Modified
  Newtonian Dynamics}.
\newblock \emph{Phys. Rev. Lett.}, 117\penalty0 (7):\penalty0 071103, 2016.
\newblock \doi{10.1103/PhysRevLett.117.071103}.

\bibitem[Ko(2016)]{Ko2016}
Chung-Ming Ko.
\newblock {On the problem of deformed spherical systems in Modified Newtonian
  Dynamics}.
\newblock \emph{Astrophys. J.}, 821\penalty0 (2):\penalty0 111, 2016.
\newblock \doi{10.3847/0004-637X/821/2/111}.

\bibitem[Verlinde(2016)]{Verlinde2016}
Erik~P. Verlinde.
\newblock {Emergent Gravity and the Dark Universe}.
\newblock 2016.

\bibitem[Bekenstein and Milgrom(1984)]{BekensteinMilgrom1984}
J.~Bekenstein and Mordehai Milgrom.
\newblock {Does the missing mass problem signal the breakdown of Newtonian
  gravity?}
\newblock \emph{Astrophys. J.}, 286:\penalty0 7--14, 1984.
\newblock \doi{10.1086/162570}.

\bibitem[Bekenstein(2004)]{Bekenstein2004}
Jacob~D. Bekenstein.
\newblock {Relativistic gravitation theory for the MOND paradigm}.
\newblock \emph{Phys. Rev.}, D70:\penalty0 083509, 2004.
\newblock \doi{10.1103/PhysRevD.70.083509, 10.1103/PhysRevD.71.069901}.
\newblock [Erratum: Phys. Rev.D71,069901(2005)].

\bibitem[Navarro and Van~Acoleyen(2006)]{NavarroVanAcoleyen2005}
Ignacio Navarro and Karel Van~Acoleyen.
\newblock {Modified gravity, dark energy and MOND}.
\newblock \emph{JCAP}, 0609:\penalty0 006, 2006.
\newblock \doi{10.1088/1475-7516/2006/09/006}.

\bibitem[Sanders(2007)]{Sanders2007}
R.~H. Sanders.
\newblock {Modified gravity without dark matter}.
\newblock \emph{Lect. Notes Phys.}, 720:\penalty0 375--402, 2007.
\newblock \doi{10.1007/978-3-540-71013-4_13}.

\bibitem[Zlosnik et~al.(2007)Zlosnik, Ferreira, and
  Starkman]{ZlosnikFerreiraStarkman2006}
T.~G Zlosnik, P.~G Ferreira, and G.~D Starkman.
\newblock {Modifying gravity with the Aether: An alternative to Dark Matter}.
\newblock \emph{Phys. Rev.}, D75:\penalty0 044017, 2007.
\newblock \doi{10.1103/PhysRevD.75.044017}.

\bibitem[Milgrom(2009)]{Milgrom2009}
Mordehai Milgrom.
\newblock {Bimetric MOND gravity}.
\newblock \emph{Phys. Rev.}, D80:\penalty0 123536, 2009.
\newblock \doi{10.1103/PhysRevD.80.123536}.

\bibitem[Blanchet and Marsat(2011)]{BlanchetMarsat2011}
Luc Blanchet and Sylvain Marsat.
\newblock {Modified gravity approach based on a preferred time foliation}.
\newblock \emph{Phys. Rev.}, D84:\penalty0 044056, 2011.
\newblock \doi{10.1103/PhysRevD.84.044056}.

\bibitem[Deffayet et~al.(2011{\natexlab{a}})Deffayet, Esposito-Farese, and
  Woodard]{DeffayetEsposito-FareseWoodard2011}
Cedric Deffayet, Gilles Esposito-Farese, and Richard~P. Woodard.
\newblock {Nonlocal metric formulations of MOND with sufficient lensing}.
\newblock \emph{Phys. Rev.}, D84:\penalty0 124054, 2011{\natexlab{a}}.
\newblock \doi{10.1103/PhysRevD.84.124054}.

\bibitem[Woodard(2015)]{Woodard2014}
R.~P. Woodard.
\newblock {Nonlocal metric realizations of MOND}.
\newblock \emph{Can. J. Phys.}, 93\penalty0 (2):\penalty0 242--249, 2015.
\newblock \doi{10.1139/cjp-2014-0156}.

\bibitem[Khoury(2015)]{Khoury2014}
Justin Khoury.
\newblock {An Alternative to particle dark matter}.
\newblock \emph{Phys. Rev.}, D91\penalty0 (2):\penalty0 024022, 2015.
\newblock \doi{10.1103/PhysRevD.91.024022}.

\bibitem[Kim et~al.(2016)Kim, Rahat, Sayeb, Tan, Woodard, and Xu]{KimEtAl2016}
M.~Kim, M.~H. Rahat, M.~Sayeb, L.~Tan, R.~P. Woodard, and B.~Xu.
\newblock {Determining Cosmology for a Nonlocal Realization of MOND}.
\newblock \emph{Phys. Rev.}, D94\penalty0 (10):\penalty0 104009, 2016.
\newblock \doi{10.1103/PhysRevD.94.104009}.

\bibitem[Blanchet and Le~Tiec(2008)]{BlanchetLeTiec2008}
Luc Blanchet and Alexandre Le~Tiec.
\newblock {Model of Dark Matter and Dark Energy Based on Gravitational
  Polarization}.
\newblock \emph{Phys. Rev.}, D78:\penalty0 024031, 2008.
\newblock \doi{10.1103/PhysRevD.78.024031}.

\bibitem[Berezhiani and Khoury(2016)]{BerezhianiKhoury20151}
Lasha Berezhiani and Justin Khoury.
\newblock {Dark Matter Superfluidity and Galactic Dynamics}.
\newblock \emph{Phys. Lett.}, B753:\penalty0 639--643, 2016.
\newblock \doi{10.1016/j.physletb.2015.12.054}.

\bibitem[Berezhiani and Khoury(2015)]{BerezhianiKhoury20152}
Lasha Berezhiani and Justin Khoury.
\newblock {Theory of dark matter superfluidity}.
\newblock \emph{Phys. Rev.}, D92:\penalty0 103510, 2015.
\newblock \doi{10.1103/PhysRevD.92.103510}.

\bibitem[Babichev et~al.(2011)Babichev, Deffayet, and
  Esposito-Farese]{BabichevDeffayetEsposito-Farese2011}
Eugeny Babichev, Cedric Deffayet, and Gilles Esposito-Farese.
\newblock {Improving relativistic MOND with Galileon k-mouflage}.
\newblock \emph{Phys. Rev.}, D84:\penalty0 061502, 2011.
\newblock \doi{10.1103/PhysRevD.84.061502}.

\bibitem[Skordis(2009)]{Skordis2009}
Constantinos Skordis.
\newblock {The Tensor-Vector-Scalar theory and its cosmology}.
\newblock \emph{Class. Quant. Grav.}, 26:\penalty0 143001, 2009.
\newblock \doi{10.1088/0264-9381/26/14/143001}.

\bibitem[Deffayet et~al.(2011{\natexlab{b}})Deffayet, Gao, Steer, and
  Zahariade]{DeffayetEtAl2011}
C.~Deffayet, Xian Gao, D.~A. Steer, and G.~Zahariade.
\newblock {From k-essence to generalised Galileons}.
\newblock \emph{Phys. Rev.}, D84:\penalty0 064039, 2011{\natexlab{b}}.
\newblock \doi{10.1103/PhysRevD.84.064039}.

\bibitem[Vainshtein(1972)]{Vainshtein1972}
A.~I. Vainshtein.
\newblock {To the problem of nonvanishing gravitation mass}.
\newblock \emph{Phys. Lett.}, B39:\penalty0 393--394, 1972.
\newblock \doi{10.1016/0370-2693(72)90147-5}.

\bibitem[Burrage et~al.(2016)Burrage, Copeland, and
  Millington]{BurrageCopelandMillington2016}
Clare Burrage, Edmund~J. Copeland, and Peter Millington.
\newblock {Radial acceleration relation from screening of fifth forces}.
\newblock 2016.

\bibitem[Hinterbichler and Khoury(2010)]{HinterbichlerKhoury2010}
Kurt Hinterbichler and Justin Khoury.
\newblock {Symmetron Fields: Screening Long-Range Forces Through Local Symmetry
  Restoration}.
\newblock \emph{Phys. Rev. Lett.}, 104:\penalty0 231301, 2010.
\newblock \doi{10.1103/PhysRevLett.104.231301}.

\bibitem[Jacobson and Mattingly(2001)]{JacobsonMattingly2000}
Ted Jacobson and David Mattingly.
\newblock {Gravity with a dynamical preferred frame}.
\newblock \emph{Phys. Rev.}, D64:\penalty0 024028, 2001.
\newblock \doi{10.1103/PhysRevD.64.024028}.

\bibitem[Bruneton(2007)]{Bruneton2006}
Jean-Philippe Bruneton.
\newblock {On causality and superluminal behavior in classical field theories:
  Applications to k-essence theories and MOND-like theories of gravity}.
\newblock \emph{Phys. Rev.}, D75:\penalty0 085013, 2007.
\newblock \doi{10.1103/PhysRevD.75.085013}.

\bibitem[Charmousis et~al.(2012)Charmousis, Copeland, Padilla, and
  Saffin]{Charmousisetal2011}
Christos Charmousis, Edmund~J. Copeland, Antonio Padilla, and Paul~M. Saffin.
\newblock {General second order scalar-tensor theory, self tuning, and the Fab
  Four}.
\newblock \emph{Phys. Rev. Lett.}, 108:\penalty0 051101, 2012.
\newblock \doi{10.1103/PhysRevLett.108.051101}.

\bibitem[Bekenstein and Sagi(2008)]{BekensteinSagi2008}
Jacob~D. Bekenstein and Eva Sagi.
\newblock {Do Newton's $G$ and Milgrom's $a_{0}$ vary with cosmological epoch
  ?}
\newblock \emph{Phys. Rev.}, D77:\penalty0 103512, 2008.
\newblock \doi{10.1103/PhysRevD.77.103512}.

\bibitem[Avilez-Lopez et~al.(2015)Avilez-Lopez, Padilla, Saffin, and
  Skordis]{AvilezEtAl2015}
Ana Avilez-Lopez, Antonio Padilla, Paul~M. Saffin, and Constantinos Skordis.
\newblock {The Parametrized Post-Newtonian-Vainshteinian Formalism}.
\newblock \emph{JCAP}, 1506\penalty0 (06):\penalty0 044, 2015.
\newblock \doi{10.1088/1475-7516/2015/06/044}.

\bibitem[Sanders(2006)]{Sanders2006}
R.~H. Sanders.
\newblock {Solar system constraints on multi-field theories of modified
  dynamics}.
\newblock \emph{Mon. Not. Roy. Astron. Soc.}, 370:\penalty0 1519--1528, 2006.
\newblock \doi{10.1111/j.1365-2966.2006.10583.x}.

\bibitem[Foster and Jacobson(2006)]{FosterJacobson2005}
Brendan~Z. Foster and Ted Jacobson.
\newblock {Post-Newtonian parameters and constraints on Einstein-aether
  theory}.
\newblock \emph{Phys. Rev.}, D73:\penalty0 064015, 2006.
\newblock \doi{10.1103/PhysRevD.73.064015}.

\bibitem[Skordis(2008)]{Skordis2008}
Constantinos Skordis.
\newblock {Generalizing tensor-vector-scalar cosmology}.
\newblock \emph{Phys. Rev.}, D77:\penalty0 123502, 2008.
\newblock \doi{10.1103/PhysRevD.77.123502}.

\bibitem[Bourliot et~al.(2007)Bourliot, Ferreira, Mota, and
  Skordis]{BourliotEtAl2006}
F.~Bourliot, P.~G. Ferreira, D.~F. Mota, and C.~Skordis.
\newblock {The cosmological behavior of Bekenstein's modified theory of
  gravity}.
\newblock \emph{Phys. Rev.}, D75:\penalty0 063508, 2007.
\newblock \doi{10.1103/PhysRevD.75.063508}.

\bibitem[Skordis et~al.(2006)Skordis, Mota, Ferreira, and
  Boehm]{SkordisEtAl2005}
Constantinos Skordis, D.~F. Mota, P.~G. Ferreira, and C.~Boehm.
\newblock {Large Scale Structure in Bekenstein's theory of relativistic
  Modified Newtonian Dynamics}.
\newblock \emph{Phys. Rev. Lett.}, 96:\penalty0 011301, 2006.
\newblock \doi{10.1103/PhysRevLett.96.011301}.

\bibitem[Carroll and Lim(2004)]{CarrollLim2004}
Sean~M. Carroll and Eugene~A. Lim.
\newblock {Lorentz-violating vector fields slow the universe down}.
\newblock \emph{Phys. Rev.}, D70:\penalty0 123525, 2004.
\newblock \doi{10.1103/PhysRevD.70.123525}.

\bibitem[{Avilez} and {Skordis}(2014)]{AvilezSkordis2014}
A.~{Avilez} and C.~{Skordis}.
\newblock {Cosmological Constraints on Brans-Dicke Theory}.
\newblock \emph{Physical Review Letters}, 113\penalty0 (1):\penalty0 011101,
  July 2014.
\newblock \doi{10.1103/PhysRevLett.113.011101}.

\bibitem[Skordis(2006)]{Skordis2005}
Constantinos Skordis.
\newblock {Teves cosmology : covariant formalism for the background evolution
  and linear perturbation theory}.
\newblock \emph{Phys. Rev.}, D74:\penalty0 103513, 2006.
\newblock \doi{10.1103/PhysRevD.74.103513}.

\end{thebibliography}

\end{document}